\pdfoutput=1
\documentclass[cernpreprint, PAPER, texlive=2020, UKenglish, texmf, orcidlogo]{atlasdoc}
 
\usepackage{atlaspackage}
\usepackage{atlasbiblatex}
 
\usepackage{atlasphysics}
 
\graphicspath{{logos/}{figures/}}
 
\usepackage{ANA-HIGG-2022-04-PAPER-defs}
\usepackage{xcolor}
\usepackage{multirow}

\AtlasTitle{Measurement of the total and differential Higgs
boson production cross-sections at $\sqrt{s} = 13$ TeV with the ATLAS detector
by combining the $H \rightarrow ZZ^* \rightarrow 4\ell$ and
$H \rightarrow \gamma \gamma$ decay channels}
 
\AtlasAbstract{
The total and differential Higgs boson production cross-sections
are measured through a combined statistical analysis of the
$H \rightarrow ZZ^* \rightarrow 4\ell$ and $H \rightarrow \gamma \gamma$
decay channels.
The results are based on a dataset of 139~$\textrm{fb}^{-1}$
of proton--proton collisions at a centre-of-mass energy of 13 TeV,
recorded by the ATLAS detector at the Large Hadron Collider.
The measured total Higgs boson production
cross-section is 55.5$^{+4.0}_{-3.8}$~pb, consistent with
the Standard Model prediction of 55.6$\pm 2.5$~pb.
All results from the two decay channels are compatible with
each other, and their combination agrees with the Standard Model
predictions.
A combined statistical interpretation of the measured fiducial
cross-sections as a function of the Higgs boson transverse momentum is
performed in order to probe the Yukawa couplings to the
bottom and charm quarks. A similar interpretation is performed by
including also the constraints from the measurements of Higgs boson
production in association with a $W$ or $Z$ boson in the
$H \rightarrow b\bar{b}$ and $c\bar{c}$ decay channels.
}
 
\author{The ATLAS Collaboration}
 
\AtlasRefCode{HIGG-2022-04}
 
\PreprintIdNumber{CERN-EP-2022-143}

\arXivId{2207.08615}

\AtlasJournal{JHEP}
\AtlasJournalRef{JHEP 05 (2023) 028}
\AtlasDOI{10.1007/JHEP05(2023)028}

\hypersetup{pdftitle={ATLAS document},pdfauthor={The ATLAS Collaboration}}
 
\begin{document}
 
\maketitle
 
\tableofcontents

\section{Introduction}
 
Following the discovery of a Higgs boson ($H$) with a mass
around 125~GeV ten years ago~\cite{HIGG-2012-27,CMS-HIG-12-028},
by the ATLAS and CMS collaborations at the
Large Hadron Collider (LHC) at CERN,
an intense programme to measure the properties of this particle and compare
them with those of the Higgs boson predicted by the Standard Model
(SM) of particle physics~\cite{Englert:1964et,Higgs:1964ia} has been
carried out.
 
In particular, total and differential fiducial Higgs
boson production cross-sections have been measured, probing the
kinematic features of the Higgs boson and of the particles
produced in association with it.
Both the ATLAS and CMS collaborations have measured
total and differential fiducial Higgs boson production
cross-sections at a proton--proton ($pp$)
centre-of-mass energy $\sqrt{s}=13$~TeV in the
$\hfourl$ (where $\ell=e,\,\mu$)~\cite{HIGG-2018-29,CMS-HIG-19-001},
$\hgg$~\cite{ATLASgamgam,CMS-HIG-17-025},
$H\rightarrow WW^{*}\rightarrow e\nu\mu\nu$~\cite{CMS-HIG-19-002},
and $H\to\tau\tau$~\cite{CMS-HIG-20-015}
decay channels.
The collaborations have also performed combinations of some of the most
sensitive results~\cite{HIGG-2017-11,CMS-HIG-17-028}.
The measurements are performed in fiducial phase spaces that closely match
the selection requirements of the detector-level analysis after the
event reconstruction.
This approach significantly reduces the model dependence that would
otherwise be introduced by relying on the acceptance factors predicted by the
model under consideration to extrapolate the measured signal yields to the
full phase space.
 
The most recent measurements of these cross-sections published by the ATLAS
collaboration, exploiting 139~\ifb~\cite{ATLAS-CONF-2019-021,LUCID2} of
13~\TeV\ proton--proton collisions produced during the whole
second data-taking phase of the LHC (Run~2,
2015--2018) and recorded by the ATLAS detector~\cite{PERF-2007-01},
have been performed using the \hfourl~\cite{HIGG-2018-29} and
\hgg~\cite{ATLASgamgam} final states.
 
The results of these two publications are combined in this article.
The measurements are extrapolated to the full phase space and the
measured cross-sections are compared with SM predictions.
Additional systematic uncertainties introduced by the
extrapolation to the full phase space are counterbalanced
by a significant reduction of the statistical uncertainty of
the measurement, which is the main limitation to the precision
of the measurements in the individual decay channels.
 
The measurements include the total production
cross-section and one and two-dimensional differential production
cross-sections as a function of the Higgs boson transverse
momentum\footnote{ATLAS uses a right-handed coordinate
system with its origin at the nominal interaction point (IP) in the
centre of the detector and the $z$-axis along the beam pipe. The
$x$-axis points from the IP to the centre of the LHC ring, and the
$y$-axis points upward. Cylindrical coordinates $(r,\phi)$ are used in
the transverse plane, $\phi$ being the azimuthal angle around the
$z$-axis. The pseudorapidity is defined in terms of the polar angle
$\theta$ as $\eta=-\ln\tan(\theta/2)$.
The rapidity of a particle of energy $E$ and longitudinal momentum
$p_z$ is defined as $y=\frac{1}{2}\ln\left(\frac{E+p_z}{E-p_z}\right)$.}
\ptH, sensitive to perturbative QCD calculations, and of the Higgs boson
rapidity \yH, sensitive to the parton distribution functions
(PDF).
Furthermore, differential cross-sections for jet multiplicity
\njets\, and the transverse momentum of the highest-$p_\mathrm{T}$ jet
\jetpt\ are also measured.
Both \njets\ and \jetpt\ observables probe the theoretical modelling of
high-$p_\mathrm{T}$ QCD radiation in Higgs boson production.
These distributions are also sensitive to the different Higgs
boson production processes.
The measurements provide a stringent test of the SM predictions and
any deviations from these predictions can indicate the presence of
physics beyond the SM (BSM).
 
This article also presents a
combined statistical
interpretation, in terms of the $b$-
and $c$-quark Yukawa coupling strengths to the Higgs boson, of the fiducial
differential cross-sections measured as a function of \ptH\ in the two decay
channels.
Another interpretation, also including the constraints on the $b$ and
$c$ Yukawa coupling strengths obtained from the measurements of
Higgs boson production in association with a $W$ or $Z$ boson, with
the Higgs boson decaying to $b$- or $c$-quark
pairs~\cite{HIGG-2018-51, ATLAS:2022ers},
is presented.
 
The results presented in this article update and supersede those of
a previous publication~\cite{HIGG-2017-11} based on the same final
states and a partial Run~2 dataset corresponding to an integrated
luminosity of 36.1~\ifb.
With respect to the previous publication, both measurements included in
this article use an improved jet reconstruction~\cite{PERF-2015-09}
and an improved unfolding procedure that is based on a detector
response matrix included in the likelihood fit.
Full descriptions of the measurements and the respective improvements in the \hfourl\ and \hgg\ decay
channels used in this article are given in
Refs.~\cite{HIGG-2018-29,ATLASgamgam}.
In both decay channels, the cross-sections in the full phase space are obtained
from these unfolded yields by taking into account the luminosity,
detector effects, acceptance factors, and branching fractions.
The SM values of the Higgs boson branching fractions are assumed, and the
acceptance factors are based on SM predictions.
The value of the Higgs boson mass is assumed to be
$125.09$~\GeV~\cite{HIGG-2014-14}.
 
The paper is organised as follows.
Section~\ref{sec:MC} describes the simulated Higgs boson event samples
and inclusive theory cross-section calculations used to obtain the
total and fiducial cross-section predictions.
The signal acceptance factors for extrapolating the results to the full phase
space are detailed in Section~\ref{sec:acceptance}.
The statistical procedure for the combination of the two channels
is illustrated in Section~\ref{sec:stat}, yielding the results
summarised in Section~\ref{sec:result}.
The differential cross-sections measured as a function of $\ptH$
are then used to constrain the Yukawa couplings of the Higgs
boson to the bottom and charm quarks in Section~\ref{sec:ly}.


\section{Higgs boson simulation samples and theoretical predictions}
\label{sec:MC}
 
The Monte Carlo (MC) event generators used for the
calculation of the acceptance factors and detector effects, and for
the SM predictions, are described in detail in
Refs.~\cite{HIGG-2018-29,ATLASgamgam}.
Their main features are summarised in this section.
 
Gluon--gluon fusion (\ggF) events are simulated using
\POWHEG NNLOPS~\cite{powheg1, powheg2, powheg3,
Hamilton:2013fea, Hamilton:2015nsa, Hamilton:2012np,
Campbell:2012am, Hamilton:2012rf, Catani:2007vq, Bozzi:2005wk,
deFlorian:2011xf}
with the \PDFforLHC[15] next-to-next-to-leading order (NNLO)
set of parton distribution functions~\cite{Butterworth:2015oua},
while other production modes are simulated with
\POWHEG~\cite{powheg1, powheg2, powheg3}
with the \PDFforLHC[15] next-to-leading order (NLO)
set except for $\bbH$ and \tH, which are
simulated using
\progname{MadGraph5\_aMC@NLO}~\cite{Alwall:2014hca, Wiesemann:2014ioa}
with the \NNPDF{3.0} NLO PDF set~\cite{Ball:2014uwa}.
These samples are generated assuming a Higgs boson with a mass \mH =
125~\gev and are normalised to cross-sections obtained from the best
available predictions as provided by the LHC Higgs Working
Group~\cite{deFlorian:2016spz} for \mH =
125.09~\gev, which are $48.5 \pm 2.4$~pb, $3.78 \pm
0.08$~pb, $2.25 \pm 0.06$~pb, $0.49 \pm 0.11$~pb and $0.59 \pm 0.05$~pb
for the \ggF, \VBF, \VH, $\bbH$ and $\ttH+\tH$ processes
respectively.
In the case of the \ggF\ NNLOPS prediction, this corresponds to a
rescaling to the fixed order $\textrm{N}^{3}\textrm{LO}$ cross-section
by a global $K$-factor of 1.1.
The impact of the 90 MeV difference between the values of \mH
used in the simulation and in the analysis
is negligible, as discussed in Section \ref{sec:acceptance}.

For all production mechanisms the \Pythia[8.2]
generator~\cite{Sjostrand:2014zea}
is used to model the \hfourl\ and \hgg\ decays, as well as for the
parton shower and the underlying event.
The \AZNLO set of tuned parameters~\cite{STDM-2012-23}
is used for \ggF, \VBF and \VH production, while the A14
tune~\cite{ATL-PHYS-PUB-2014-021} is used for the other production
modes.
Alternative \ggF, \VBF, \VH, \tH (\ttH) samples are
produced by interfacing the nominal matrix element generator
with \HERWIG[7.1.3] (\HERWIG[7.0.4])~\cite{Bahr:2008pv,Bellm:2015jjp},
using the H7UE set of tuned parameters~\cite{Bellm:2015jjp}, in order
to estimate uncertainties in the signal acceptance factors
related to the modelling of the parton shower.
 
The measurements are also compared with an alternative prediction
obtained by summing the expected cross-sections of non-\ggF Higgs boson
production processes described previously and an alternative SM
\ggF prediction obtained using
\progname{MadGraph5\_aMC@NLO} (\fxfx).
This matrix-element generator provides NLO accuracy in QCD for zero, one, and
two additional jets, using the \progname{FxFx} merging
scheme~\cite{Alwall:2014hca,Frederix:2012ps}, and includes the
top and bottom quark mass
effects~\cite{Frederix:2016cnl,Mantler:2015vba,Mantler:2012bj}.
The events are generated using the
\NNPDF{30} NLO PDF set.
The generator is interfaced to \Pythia[8] for the modelling of
the parton shower.
The predicted cross-sections are scaled
by a global $\textrm{N}^{3}\textrm{LO}$ $K$-factor of 1.47.
 
Uncertainties in the predicted \ggF, \VBF, \VH and $\ttH$
cross-sections induced by PDF uncertainties are estimated by
varying the \progname{PDF4LHC} set according to its
eigenvectors~\cite{Butterworth:2015oua}, and summing in quadrature the
variations in the predictions.
The effect of PDF variations on the $tH$ and $\bbH$ cross-sections has a
negligible impact on the total uncertainty and is not included.
 
Uncertainties due to missing higher-order QCD effects for the
\ggF~\progname{NNLOPS}, \VBF, \VH and $\ttH$ predicted cross-sections
are estimated using the same scheme as in
Refs.~\cite{HIGG-2018-29,ATLASgamgam}: parameters accounting
for cross-section and migration effects across various Higgs boson
kinematic and associated jet observables are used and their variations
are summed in quadrature.
For other production modes, uncertainties
related to missing higher-order QCD effects are estimated by
varying the renormalisation and factorisation scales by factors of
0.5 and 2.0, and computing the difference between the envelope of the
alternative predictions and the nominal one.
 
The Higgs boson branching ratios for $\mH=125.09$~\gev\ are assumed
to be those of the SM, (0.0125 $\pm$ 0.0003)\% for the four-lepton
final state and (0.227 $\pm$ 0.007)\% for the diphoton final
state~\cite{deFlorian:2016spz}.


\section{Acceptance factors}
\label{sec:acceptance}
 
The acceptance factors that extrapolate at particle-level from the
respective \hfourl\ and \hgg\ fiducial phase spaces to the full phase
space are estimated using the simulated event samples and
cross-sections described in Section~\ref{sec:MC}.
The definitions of the fiducial phase spaces
are summarised in Table~\ref{tab:H4lFiducialRegion} and
Table~\ref{tab:HggFiducialRegion}, with more details provided
in Refs.~\cite{HIGG-2018-29,ATLASgamgam} respectively.
The evaluation of the acceptance factors assumes SM Higgs boson
production fractions and a Higgs boson mass of
125~\GeV: the 90 \MeV\ difference from the measured mass value of
125.09 \GeV\ has a negligible impact on the Higgs boson kinematics.
 
\begin{table}[!htbp]
\centering
\caption{Summary of the particle-level fiducial definitions in the
\hfourl\ analysis~\cite{HIGG-2018-29}.
A lepton {\em quadruplet} is formed by two
same-flavour, opposite-charge ({\em SFOC}) lepton pairs.
Dressed leptons are leptons whose four-momenta have been
modified by adding the four-momenta of photons within a cone
of size $\Delta R = 0.1$ around the lepton to account for final
state radiation.
The invariant mass of the SFOC lepton pair that is closest to $m_Z$ is
denoted with $m_{12}$, while the invariant mass of the SFOC pair
of remaining leptons that is closest to $m_Z$ is denoted with $m_{34}$.
The quadruplet satisfying the \textit{lepton
selection and pairing criteria} is labelled as the nominal
quadruplet.
If the nominal quadruplet fails the \textit{event
selection criteria}, no quadruplet is marked as the Higgs boson
candidate. If the nominal quadruplet passes the selection and
there is an additional lepton, the quadruplet with the largest
\ggF matrix element value is taken as the Higgs boson candidate. If no
extra lepton is found, then the nominal quadruplet is taken as
the Higgs boson candidate.
}
\label{tab:H4lFiducialRegion}
\vspace{0.15cm}
\begin{tabular}{ll}
\toprule
\multicolumn{2}{c}{\textbf{Lepton and jet definitions}} \\
\midrule
Leptons & Dressed leptons not originating from hadron or $\tau$ decays  \\
& $\pt>5$ \GeV, $|\eta|<2.7$  \\
Jets &  \pt~$>$ 30 \gev, $|y|<$ \,4.4 \\
\midrule
\multicolumn{2}{c}{\textbf{Lepton selection and pairing}} \\
\midrule
Lepton kinematics & $\pt$ threshold for three leading leptons: $> 20, 15, 10 $ \GeV  \\
Leading pair ($m_{12}$) & SFOC lepton pair with smallest $|m_Z-m_{\ell\ell}|$  \\
Subleading pair ($m_{34}$) & Remaining SFOC lepton pair with  smallest $|m_Z-m_{\ell\ell}|$ as nominal \\
\midrule
\multicolumn{2}{c}{\textbf{Event selection}} \\
\midrule
Mass requirements & 50 \GeV$ < m_{12} < 106$ \GeV \, and  12 \GeV $< m_{34} < 115$ \GeV  \\
Lepton separation & $\Delta R(\ell_{i},\ell_{j}) > 0.1$ \\
Lepton/Jet separation & $\Delta R(\ell_{i}, \textrm{jet}) > 0.1$ \\
$J/\psi$~veto & $m(\ell_{i},\ell_{j}) > 5$ \GeV \,  for all SFOC lepton pairs  \\
Mass window &  105~\GeV $< m_{4\ell} < 160 $ \GeV  \\
If extra lepton with $\pt>12$ GeV&  Quadruplet with largest \ggF matrix element value \\
\bottomrule
\end{tabular}
\end{table}
 
\begin{table}[!htbp]
\centering
\caption{Summary of the particle-level fiducial definitions in the
\hgg\ analysis~\cite{ATLASgamgam}. $\ET^{\textrm{iso}}(\Delta R, p_{\textrm{T}},
\textrm{charged})$ is the scalar sum of the transverse momenta
of charged stable particles with a transverse momentum above
the specified threshold
within a $\Delta R$ cone centred on the photon
direction.}
\label{tab:HggFiducialRegion}
\begin{tabular}{l c}
\toprule
\multicolumn{2}{c}{\textbf{Photon and jet definitions}} \\
\midrule
Photons     &  Photons not originating from hadron decays \\
&  $\pt>15$ \GeV, $|\eta| < 1.37$ or $1.52 < |\eta| < 2.37$ \\
& $\ET^{\textrm{iso}}(\Delta R<0.2, p_{\textrm{T}} > 1$~\gev, charged) $< 0.05\ E_{\textrm{T}}$ \\
Jets        & $\pt > 30$ GeV, $|y| < 4.4$ \\
\midrule
\multicolumn{2}{c}{\textbf{Event selection}} \\
\midrule
Photon kinematics & $\pt$ threshold for two leading photons: $\pt^{\gamma_1} > 0.35m_{\gamma\gamma}$, $\pt^{\gamma_2} > 0.25m_{\gamma\gamma}$ \\
Mass window & $105\ \gev < m_{\gamma\gamma} < 160$~\gev \\
\bottomrule
\end{tabular}
\end{table}
 
In the full phase space, the quantities \ptH\ and \yH\ are computed
directly from the simulated Higgs boson momentum instead of its decay
products, as in the fiducial analyses. The acceptance factors implicitly
include the correction for this difference. Simulated particle-level jets
are built from all stable particles with $c\tau > 10$ mm, including
neutrinos, photons, and leptons from hadron decays or produced in the
shower. All decay products from the Higgs boson decay and
the leptonic decays of associated vector bosons are removed from the
inputs to the jet algorithm. Jets are reconstructed using the
anti-$k_t$ algorithm~\cite{Cacciari:2008gp} with a radius parameter
$R = 0.4$, and are required to have $\pT > 30$~\GeV.
 
Theory uncertainties related to the PDF, higher-order corrections, and
the parton shower model are taken into account when evaluating acceptance
factors. For each channel, the uncertainties in the acceptance factors
are correlated with the impact of these theoretical sources on the
detector response matrix used in the unfolding.
Due to this procedure, compared with the results
in Ref.~\cite{ATLASgamgam}, the \hgg\ results presented in this article
have these additional theoretical uncertainties in the detector
response matrix. Uncertainties due to the PDF and missing higher-order
corrections are estimated as described in
Section~\ref{sec:MC}.
Uncertainties due to the parton shower model are evaluated by comparing the
acceptance factors estimated using MC samples with the default
\progname{Pythia8}
showering with
the acceptance factors computed using MC samples relying on
the \progname{Herwig7}
showering model.
To account for the uncertainties in the SM Higgs boson production
cross-sections when calculating the total acceptance factor from the sum of
the various production modes,
the fractions of production modes are independently varied within their
measured uncertainties taken from Ref.~\cite{HIGG-2018-57}. The total
systematic uncertainties in the acceptance factors range between 0.5\%
and 7\%, depending on the observable and bin, with the parton shower
uncertainty being the dominant source.
 
The inclusive acceptance factors, relative to the full phase space,
are about 50\% for both the \hfourl\ and the \hgg\ channels.
Figure~\ref{fig:accFactorsSum} shows the
acceptance factors and their systematic uncertainties as a function of
\ptH\ and \njets.
In the \hfourl\ channel, the acceptance factor drops in the
highest \ptH\ bin due to the lepton
separation requirement, while the shape of the acceptance factor
for the \hgg\ channel as a function of \ptH\ is due to the $p_\mathrm{T}$
selection criteria on the photons.

\begin{figure}[!htbp]
\centering
\subfloat[]{\includegraphics[width=0.48\linewidth]{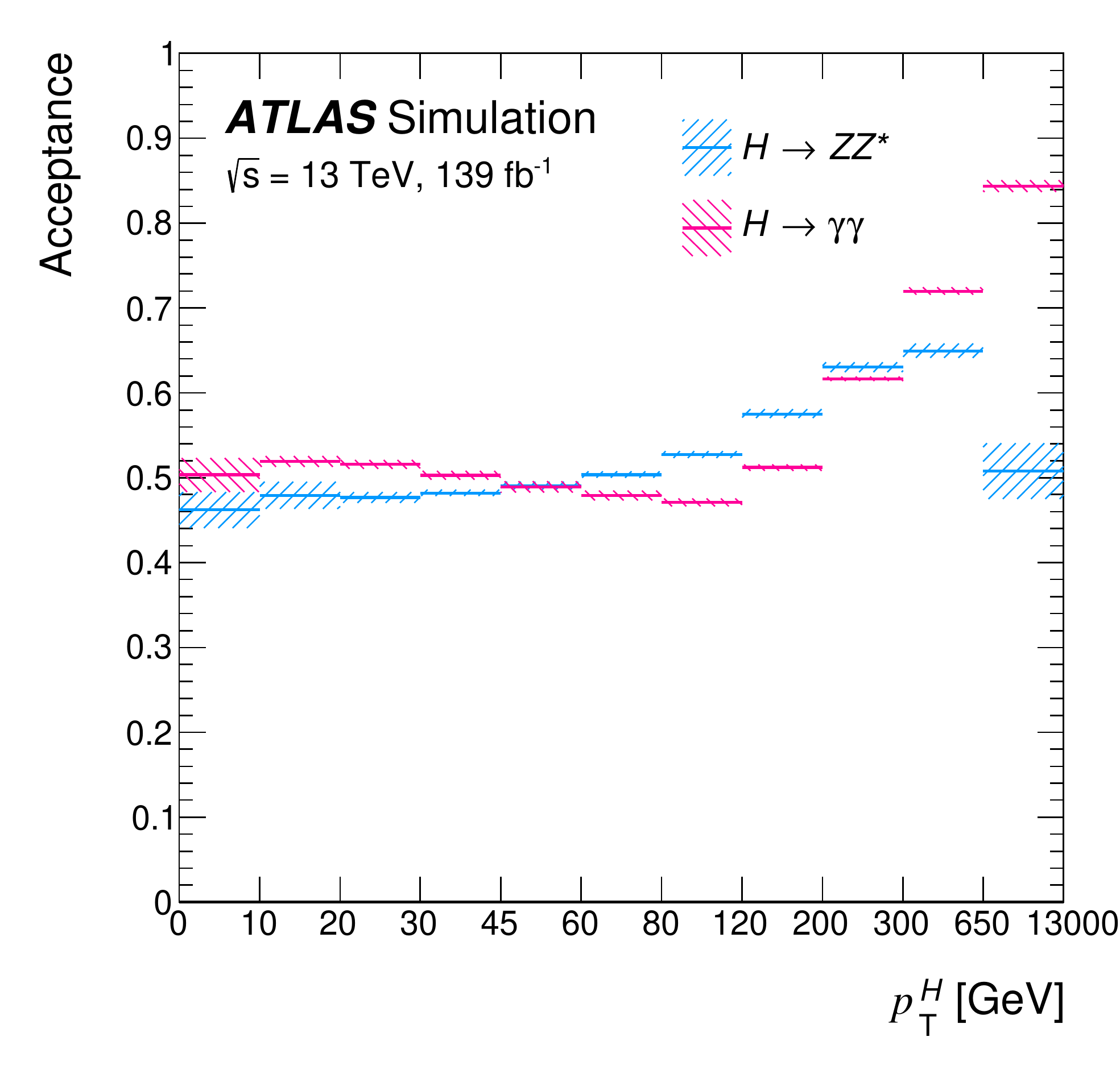} \label{fig:Acc_pt}}
\subfloat[]{\includegraphics[width=0.48\linewidth]{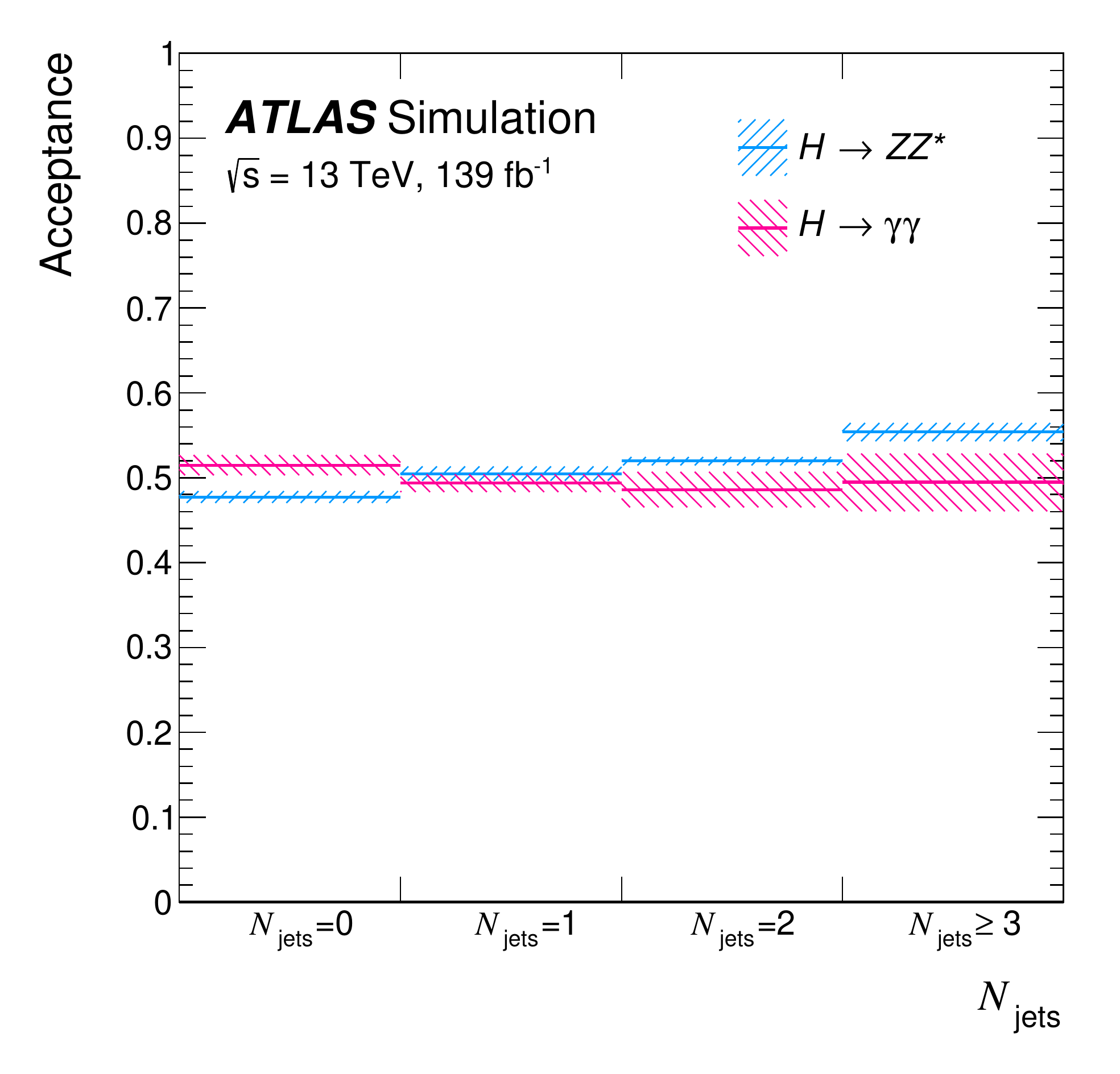}\label{fig:Acc_Njets}}
 
\caption{Acceptance factors (solid lines), including systematic
uncertainties (hatched bands), for the extrapolation from the
fiducial to the full phase space for the \hfourl\ decay channel
(blue) and the \hgg\ decay channel (magenta), as a function of variables
characterising the Higgs boson kinematics:
\protect\subref{fig:Acc_pt} Higgs boson transverse momentum
\ptH\
and \protect\subref{fig:Acc_Njets} number of jets \njets\ with
$\pT > 30$~\GeV.
}
\label{fig:accFactorsSum}
\end{figure}


\section{Statistical procedure}
\label{sec:stat}
 
A likelihood combination of the two decay channels is performed,
following the method described in Ref.~\cite{HIGG-2017-11}.
For some observables, such as \ptH\ and \jetpt, the binning in
the $\hgg$ analysis is finer than that in the $\hfourl$ analysis.
Where needed, the sum of the consecutive \hgg\ sub-bins is
combined with one \hfourl\ bin such
that the measured bin boundaries match between the two results.
A summary of the bin boundaries used in the combined results is
presented in Table \ref{tab:BinDefinition}.
Higgs boson events that are outside of the fiducial region but
are reconstructed within the signal region are accounted for
in the likelihood function by a small correction (around 1--2\%)
of the signal normalisation, as described in
Refs.~\cite{HIGG-2018-29,ATLASgamgam}.
 
\begin{table}[!h]
\centering
\caption{Bin boundaries used in the combination of the
cross-section of various differential observables. For the
\jetpt\ distribution, the first bin contains all events with a
leading jet with \pt less than 30 \GeV, and corresponds exactly
with the 0-jet bin in the \njets\ differential
distribution.  \label{tab:BinDefinition}}
\begin{tabular}{ c c c }
\toprule
Variable  & Bin Edges & $N_{\mathrm{bins}}$  \\
\midrule
\ptH   & 0, 10, 20, 30, 45, 60, 80, 120, 200, 300, 650, 13000~\gev & 11 \\
\yH    & 0, 0.15, 0.3, 0.45, 0.6, 0.75, 0.9, 1.2, 1.6, 2.0, 2.5 & 10 \\
\njets & 0, 1, 2, $\geq$ 3 & 4 \\
\jetpt & 0, 30, 60, 120, 350~\gev & 4 \\
\ptH\ vs \yH & \ptH: 0, 45, 120, 350~\gev; \yH: 0, 0.5, 1.0, 1.5, 2.5 & 12 \\
\bottomrule
\end{tabular}
 
\end{table}

Experimental and theoretical uncertainties that affect both channels
are correlated via common nuisance parameters. The correlated
experimental uncertainties include the uncertainties in the integrated
luminosity, in the description of the pile-up in the simulation, in the
jet reconstruction and calibration, in the common electron-photon
energy scale, in the Higgs boson mass value, and in the contributions
of the different  Higgs boson production modes.
Additionally, the common sources of theoretical uncertainty in the
$\hfourl$ and $\hgg$ branching ratios (strong coupling constant, $b$ and
$c$ quark masses, and partial decay widths to the main decay channels, such
as two vector bosons, two gluons, or a $b\bar{b}$ pair) are also correlated.
Finally, the theoretical uncertainties in the acceptance factor and response
matrix due to missing higher-order QCD effects, PDF variations,
variations of the modelling of the parton shower, and signal
composition uncertainties are also correlated across the Higgs
boson decay channels.
 
The asymptotic approximation~\cite{Cowan:2010js}
for the distribution of the profile likelihood ratio is assumed
in the computation of uncertainties on all reported measurements.
The validity of this approximation has been verified in previous
analyses by performing pseudo-experiments.


\section{Results}
\label{sec:result}
 
The total Higgs boson production cross-section at $13~\TeV$ is
measured to be $53.0^{+5.3}_{-5.1}$~pb
\hbox{$(^{+4.9}_{-4.8} \mathrm{(stat.)}~^{+2.0}_{-1.7} \mathrm{(syst.)})$}
using the \hfourl\ decay channel and $58.1^{+5.7}_{-5.4}$~pb
\hbox{$(\pm 4.2  \mathrm{(stat.)}~^{+3.9}_{-3.5} \mathrm{(syst.)})$}
using the \hgg\ decay channel.
The total cross-section obtained combining the two results
is $55.5^{+4.0}_{-3.8}$~pb
\hbox{$(\pm 3.2 \mathrm{(stat.)}~^{+2.4}_{-2.2} \mathrm{(syst.)})$}.
All three results are in agreement with the SM prediction of
$55.6\pm 2.5$~pb.
The measurements in the two decay channels are compatible with
each other with a $p$-value of 49\%, and the compatibility of
the combined result with the SM prediction has a $p$-value of 98\%.
All compatibility checks are performed using a likelihood ratio
approach, based on the test statistic variation under different
hypotheses in the asymptotic approximation.
 
\begin{figure}[!htbp]
\centering
\includegraphics[width=0.75\linewidth]{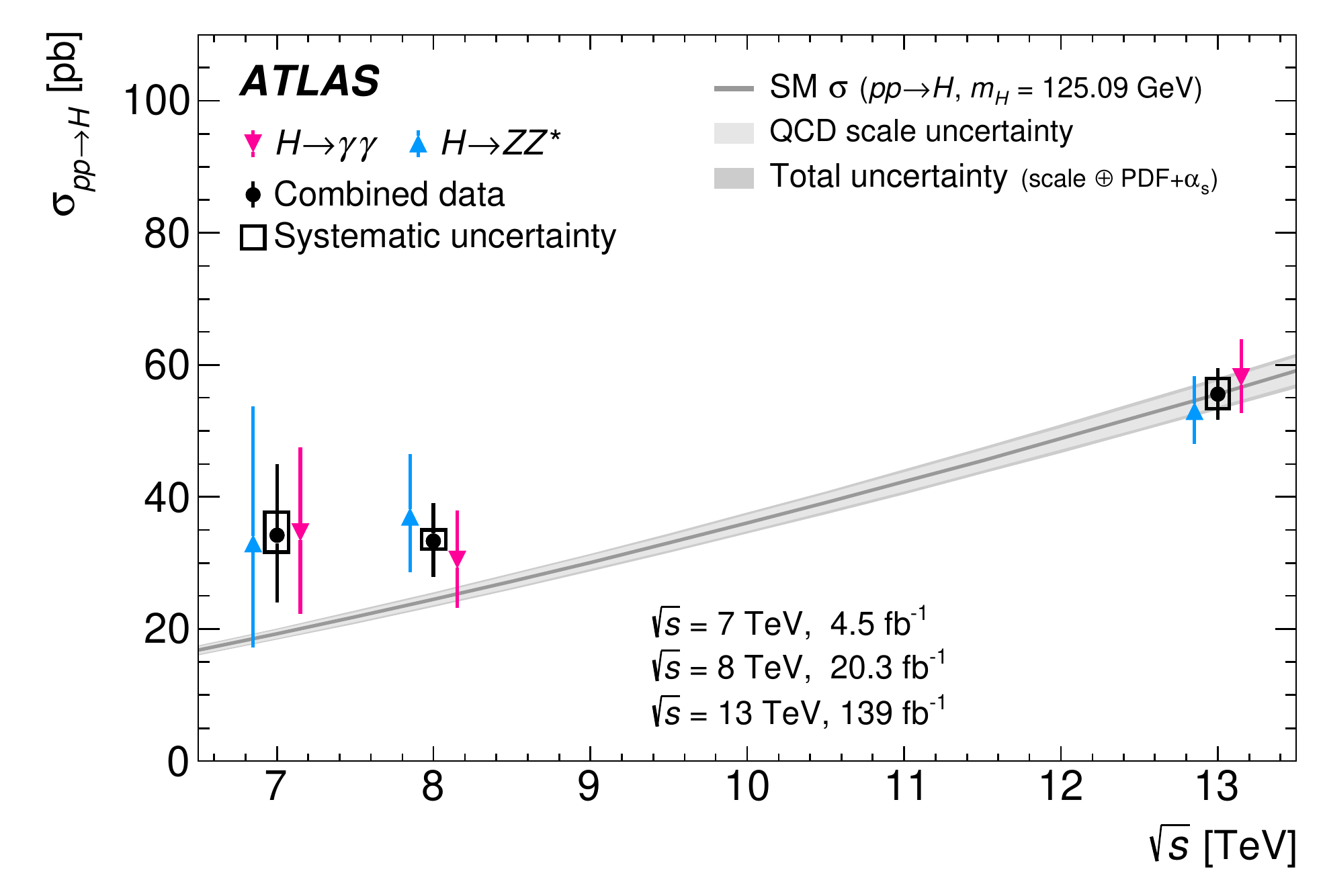}
\caption{Total $pp \rightarrow H + X$ cross-sections measured at
centre-of-mass energies of 7, 8 and
13~\TeV, compared with Standard Model predictions taken from
Ref.~\cite{deFlorian:2016spz}. The measurements with the
\hfourl\ channel (blue triangles),
\hgg\ channel (magenta inverted triangles) and their combination
(black dots) are shown. The
individual channel results are offset along the $x$-axis for
display purposes. The black boxes around the combined measurements
represent the systematic uncertainty, while the error bars show
the total uncertainty. The light grey band shows the
uncertainty in the prediction due to missing QCD higher-order
corrections. The dark grey band indicates the total
theoretical uncertainty, corresponding to the dominant
QCD higher-order-correction uncertainty summed in quadrature with
the sum of the PDF and $\alpha_S$ uncertainties, and is
partially correlated across values of the centre-of-mass
energy.}
\label{fig:totalxs_sqrts}
\end{figure}
 
The total cross-section measured using the two channels, their combination,
and the SM prediction for a Higgs boson mass of 125.09~GeV
are shown in Figure~\ref{fig:totalxs_sqrts}.
The figure also includes the results of the measurements using data
collected at a $pp$ centre-of-mass energies of
$\sqrt{s}=8$~TeV and 7~TeV, and the corresponding theoretical
expectations.
The event samples, selections and the cross-section measurement techniques
used for the 8~TeV measurements are described in
Refs.~\cite{HIGG-2013-22,HIGG-2013-10}; similar techniques are used
to measure the cross-sections at 7~TeV as
described in Refs.~\cite{HIGG-2013-21,HIGG-2013-08}.
For both the 7 and 8~TeV results, the signal yields in the two decay
channels are measured inclusively
and corrected for acceptance and detector effects.
The results at each centre-of-mass energy
are then combined using a likelihood-based
technique described in Ref.~\cite{HIGG-2014-11}.
The total Higgs boson production cross-section at $7~\TeV$ is measured
to be $33^{+21}_{-16}$~pb using the \hfourl\ channel, 35$^{+13}_{-16}$~pb
using the \hgg\ decay channel, and
$34^{+11}_{-10}$~pb \hbox{($\pm 10 \mathrm{(stat.)}~^{+4}_{-2}
\mathrm{(syst.)}$)} from their combination. This is
to be compared with the SM expectation of $19.2 \pm 0.9$~pb.
At 8~\tev, the total Higgs boson production cross-section
is measured to be 37$^{+9}_{-8}$~pb using the \hfourl\ channel,
30.5$^{+7.5}_{-7.4}$~pb using the \hgg\ decay channel,
and 33.3$^{+5.8}_{-5.4}$~pb
\hbox{($^{+5.5}_{-5.3} \mathrm{(stat.)}~^{+1.7}_{-1.3} \mathrm{(syst.)}$)}
from their combination. This is to be compared with the SM expectation of
$24.5 \pm 1.1$~pb.
These results supersede the previously published ones, which used
signal yield estimates, acceptance factors and SM Higgs boson branching
ratios calculations based on a different value of the Higgs boson mass.
 
The differential cross-sections in bins of \ptH, \yH, \ptH\ vs \yH,
\njets\ and \jetpt\ for the
individual channels and their combination are shown in
Figures~\ref{fig:diffxs} and~\ref{fig:diffxs2}.
The uncertainty band around the SM prediction includes PDF and $\alpha_S$
uncertainties as well as those due to missing QCD higher-order
corrections, obtained following the method described in
Ref.~\cite{HIGG-2016-21}. When compared with the results from the
individual channels, the total uncertainty for the combined results is
lower by 20\%--40\% and the impact of uncorrelated systematic
uncertainties is reduced by approximately 40\%.
The observed correlation matrices among the cross-sections
measured in different bins of the same observable
are shown in Appendix~\ref{app:corrmat}. The correlations are small ($<10\%$)
for the Higgs-related observables ($\ptH$, $\yH$), characterised
by better experimental resolution, and larger (up to about 40\%)
for jet-related observables (\njets\ and \jetpt) with worse
resolution and larger migrations.

All combined measurements are dominated by statistical
uncertainties.
Significant systematic uncertainties affecting the total
and all differential cross-sections
arise from the background modelling in the \hgg\ signal
extraction \cite{ATLASgamgam} (typical error of 2--5\%) and
the integrated luminosity (1.7\%).
For the \njets\ and \jetpt\ differential cross-section measurements,
the uncertainties in the reconstruction of the jet
energy scale and resolution are important as well, with impacts
on the results typically in the range of 2--9\%.
The dominant theoretical source of uncertainty is the
parton shower modelling for \ggF signal and has an impact of 2--6\%.
 
\begin{figure}[!htbp]
\centering
\subfloat[]{\includegraphics[width=0.48\linewidth]{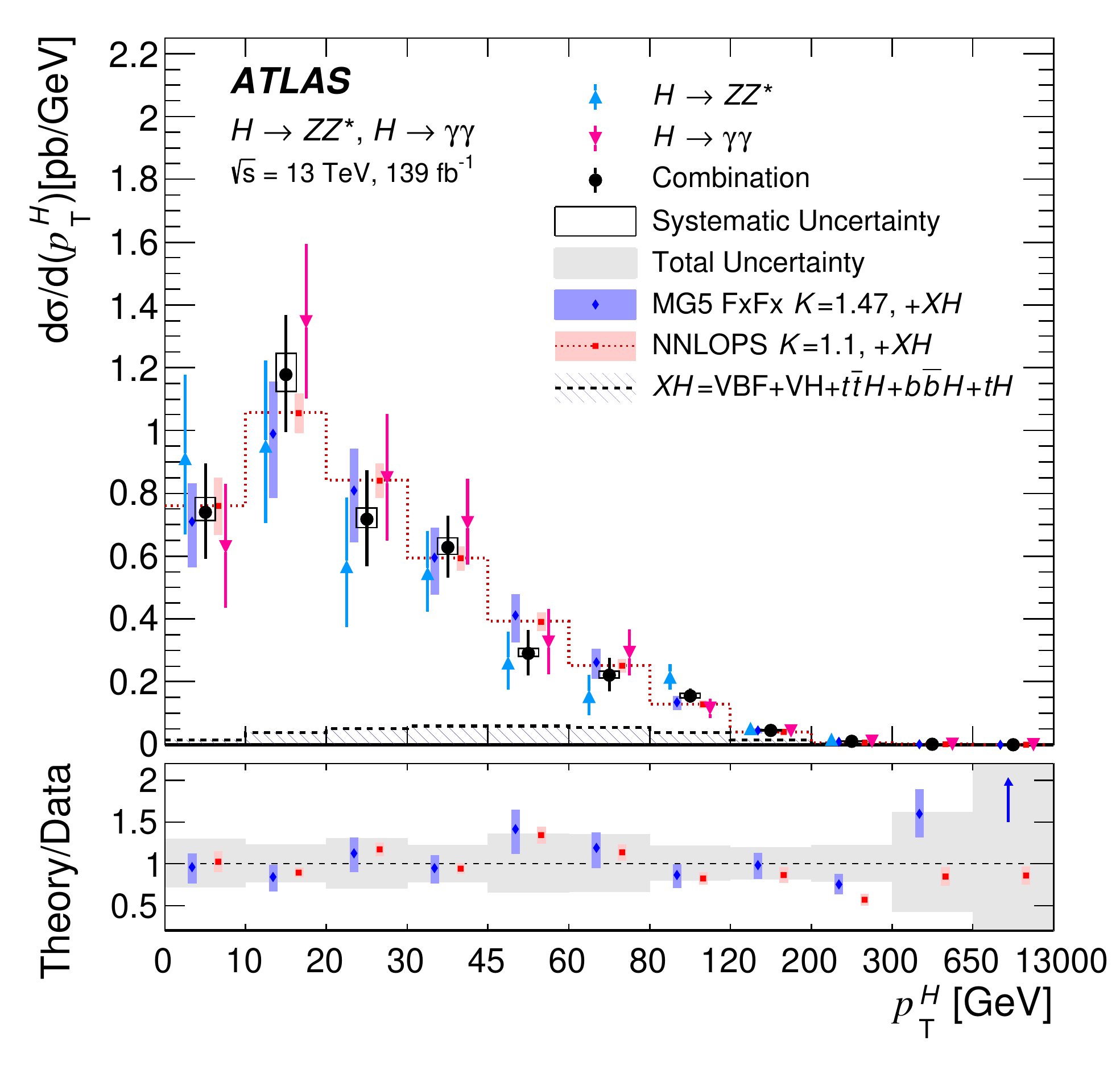}\label{fig:summary_pt}}
\subfloat[]{\includegraphics[width=0.48\linewidth]{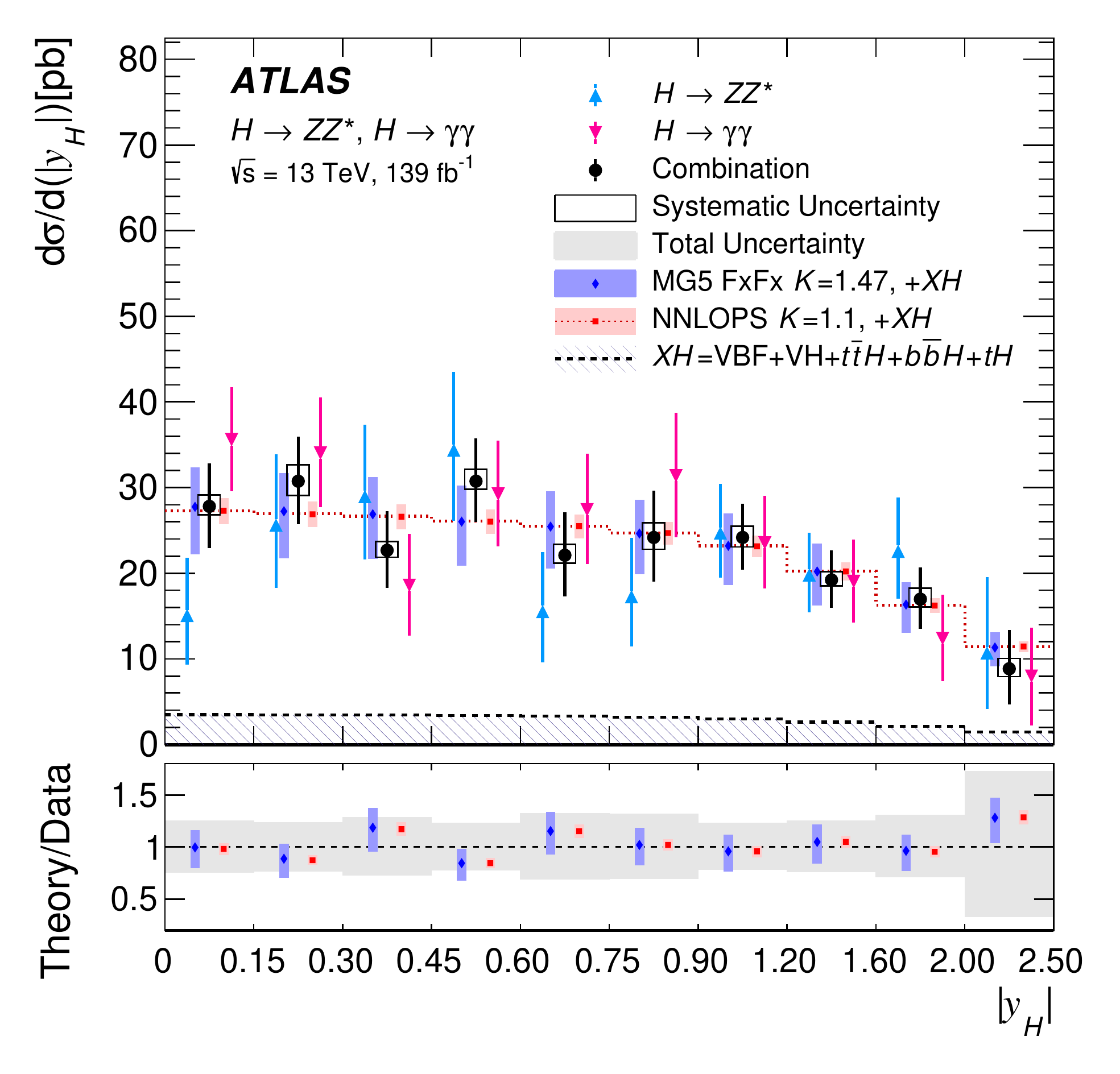}\label{fig:summary_y}}\\
\subfloat[]{\includegraphics[width=0.48\linewidth]{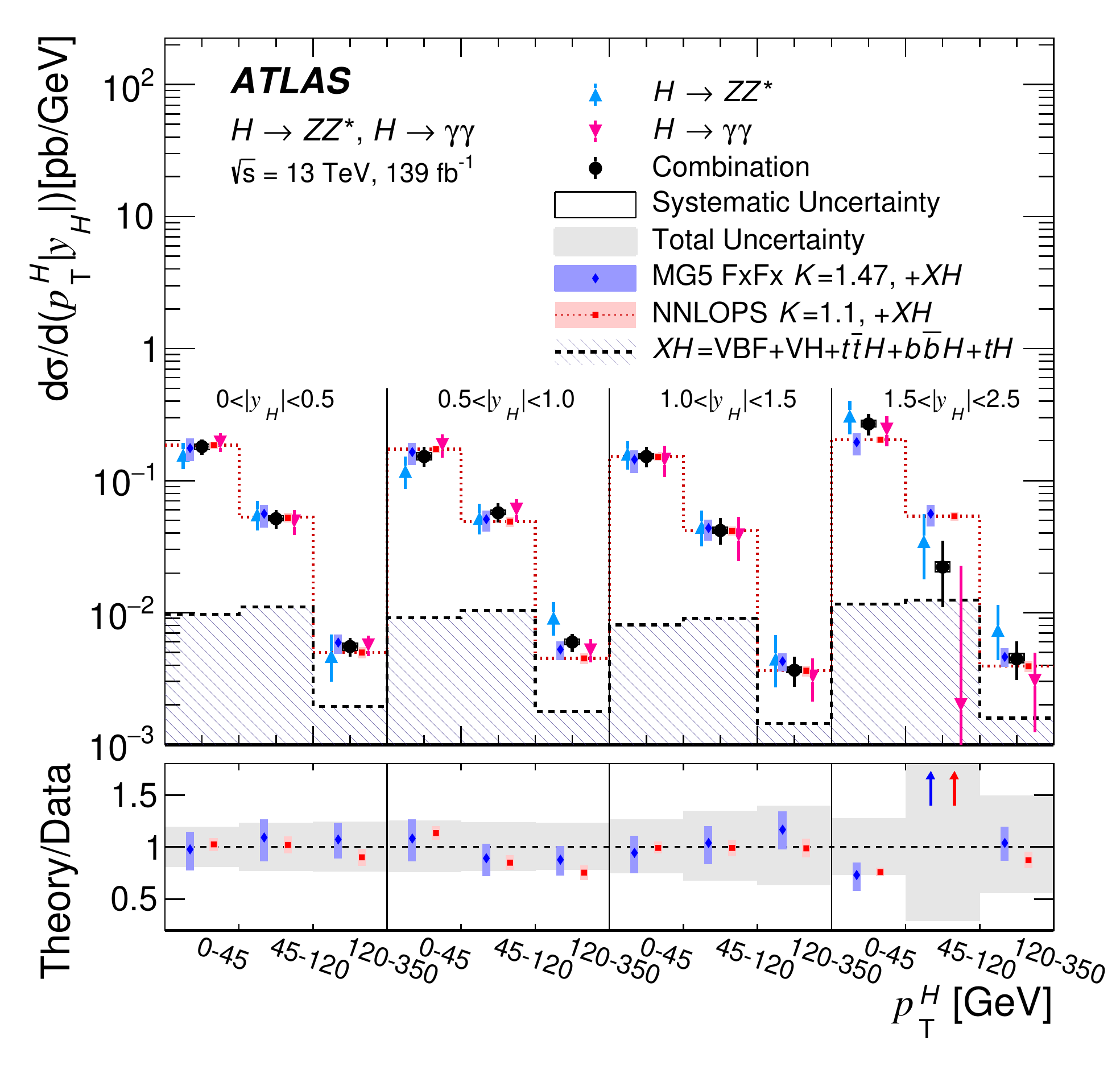}\label{fig:summary_pt_vs_y}}
\caption{Differential $pp \rightarrow H + X$ cross-sections,
in the full phase space, as a function of variables characterising
the Higgs boson kinematics:
\protect\subref{fig:summary_pt} Higgs boson transverse
momentum \ptH, \protect\subref{fig:summary_y} Higgs boson
rapidity \yH, and
\protect\subref{fig:summary_pt_vs_y} \ptH\ vs \yH,
compared with Standard Model predictions.
The $\hfourl$ (blue triangles), $\hgg$ (magenta inverted triangles),
and combined (black squares) measurements are shown.
The error bars on the data points show the total uncertainties, while
the systematic uncertainties are indicated by the boxes.
The measurements are compared with two predictions, obtained by
summing the \ggF predictions of \NNLOPS or \fxfx,
normalised to the fixed order N$^3$LO total cross-section with the
listed $K$-factors, and the
MC predictions for the other production processes $XH$.
The shaded bands indicate the relative impact of the PDF
and scale systematic uncertainties in the prediction. These
include the uncertainties related to the $XH$ production modes.
The dotted red histogram corresponds to the central value of the
prediction that uses \NNLOPS for the modelling of the \ggF\ component.
The bottom panels show the ratios between the predictions
and the combined measurement.
The grey area represents the total uncertainty of the measurement.
For better visibility, all bins are shown as having the same size,
independent of their numerical width.}
\label{fig:diffxs}
\end{figure}
 
\begin{figure}[!htbp]
\centering
\subfloat[]{\includegraphics[width=0.48\linewidth]{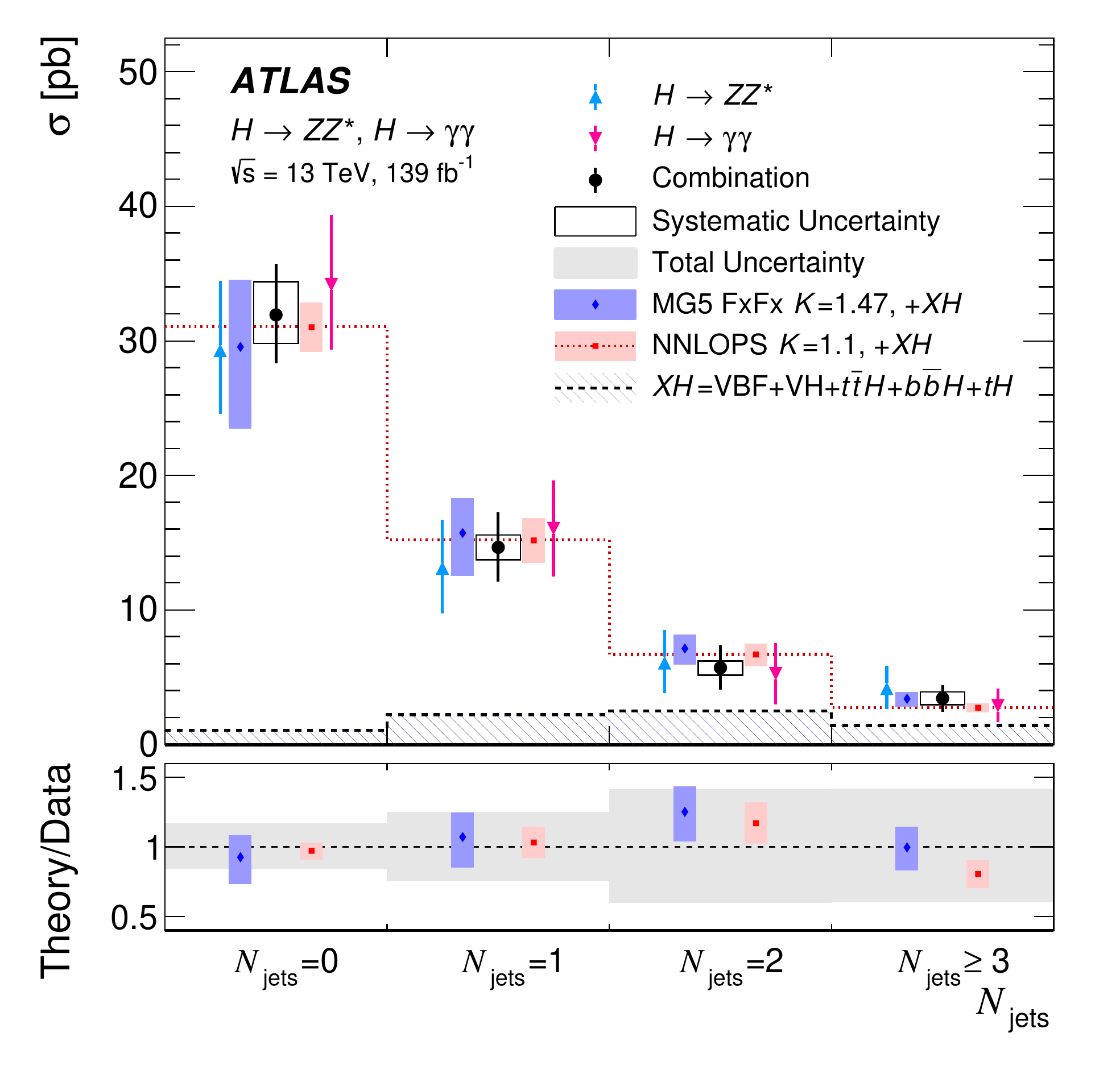}\label{fig:summary_Njet}}
\subfloat[]{\includegraphics[width=0.48\linewidth]{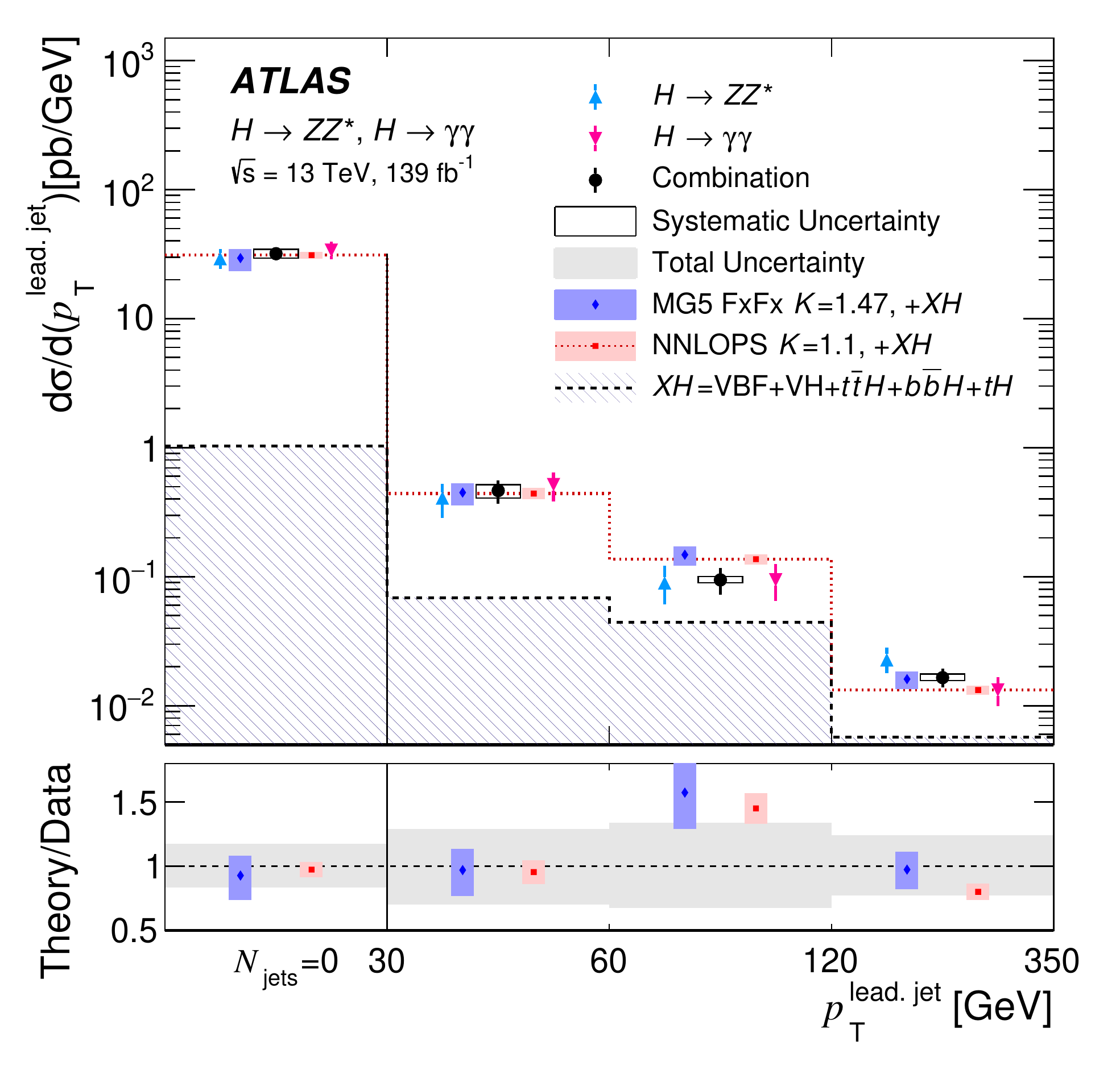}\label{fig:summary_leadjet}}
\caption{Differential $pp \rightarrow H + X$ cross-sections,
in the full phase space, as a
function of variables related to the jets
produced in association with the Higgs boson,
\protect\subref{fig:summary_Njet} number of jets
and \protect\subref{fig:summary_leadjet} $\pT$ of the leading jet,
compared with Standard Model predictions.
The figure uses the same layout as Figure~\ref{fig:diffxs}.
}
\label{fig:diffxs2}
\end{figure}
 
The $p$-values for the compatibility among the individual
measurements are given in Table~\ref{tab:SummaryCompatibility}.
The $p$-values for the compatibility of the measurements with
various theoretical predictions are given in
Table~\ref{tab:SummaryDiffXSPval} for the differential cross-section
results.
For all observables, the measurements in the two channels are compatible
with each other, with $p$-values ranging between 20\% and 80\%.
The combined measurements are also in good agreement with the predictions,
with $p$-values ranging between 20\% and 98\%. The prediction based
on the \progname{NNLOPS} simulation of gluon--gluon fusion events is
lightly favoured over that based on the \fxfx\ simulation.

\begin{table}[!htbp]
\centering
\caption{$p$-values for the compatibility of the
individual \hfourl\ and \hgg\ results for the combined total and
differential cross-sections.}
\label{tab:SummaryCompatibility}
\vspace{0.2cm}
\footnotesize
\renewcommand{\arraystretch}{1.2}
\begin{tabular}{ l c c c c c c}
\toprule
Observable    & Total & \ptH  & \yH   & \ptH\ vs \yH & \njets & \jetpt \\
\midrule
Compatibility $p$-value &  49\%     &  20\% &  23\% & 69\%        &  80\%  &  37\%  \\
\bottomrule
\end{tabular}
\end{table}
 
\begin{table}[!htbp]
\centering
\caption{$p$-values for the compatibility of the
measured cross-sections with the SM predictions when the
distributions for gluon--gluon fusion events obtained with either
\progname{NNLOPS} or
\progname{\fxfx},
scaled to the fixed order N$^3$LO total gluon--gluon fusion cross-section,
are used.
The uncertainties in the theoretical predictions are included when
calculating the $p$-values.}
\label{tab:SummaryDiffXSPval}
\vspace{0.2cm}
\footnotesize
\renewcommand{\arraystretch}{1.2}
\begin{tabular}{ l c c c c c c }
\toprule
SM prediction         &   \ptH  & \yH   & \ptH\ vs \yH   & \njets & \jetpt  \\
\midrule
\progname{NNLOPS}     &   91\%  & 98\%  & 56\%  & 95\%   &  34\%   \\
\progname{\fxfx}      &   73\%  & 98\%  & 56\%  & 86\%   &  23\%   \\
\bottomrule
\end{tabular}
\end{table}
 
\clearpage


\section{Constraints on the $b$- and $c$-quark Yukawa couplings}
\label{sec:ly}
 
The observations of the Higgs boson decays to
$b\bar{b}$~\cite{HIGG-2018-51,CMS-HIG-18-016} provided stringent
constraints on the possible modification of the $b$-quark Yukawa
coupling with respect to its SM prediction,
whereas current searches for Higgs boson decays to charm final
states~\cite{ATLAS:2022ers,CMS-HIG-18-031} still allow for
a relatively large modification of the \cquark coupling.
These measurements have been interpreted in terms of the Yukawa coupling
modifiers for \(b\)- and \(c\)-quarks, \kappab and \kappac,
defined as multipliers of the SM values of these
couplings~\cite{deFlorian:2016spz}.
The measured value of $\kappab$ agrees with the SM prediction
of one with a precision of about 10\%~\cite{HIGG-2021-23}
to 20\%~\cite{CMS-HIG-22-001},
whereas the constraints on \kappac are significantly looser:
$|\kappa_c|<5.7$~\cite{HIGG-2021-23}
or $1.1<|\kappa_c|<5.5$~\cite{CMS:2022psv}
at the 95\% confidence level (CL).
 
The Higgs boson \pt distribution is sensitive to modifications of
the Yukawa couplings of the Higgs boson to the $b$-
and $c$-quarks~\cite{Bishara:2016jga}.
This sensitivity is driven by the contributions of $b$- and $c$-quarks
to the loop-induced \ggF production and by the quark-initiated
production of the Higgs boson.
The former production mode includes an interference term between
$b$- and $c$-quark loop-mediated amplitudes which is proportional
to the product of the two couplings and is therefore
sensitive to their relative sign.
Modifications of the coupling strength to $b$- and \cquarks
result in changes to both the
overall cross-section and the shape of the \ptH\ distribution.
In addition, the branching ratio for the \hgg\ decay would be affected
by corresponding changes to its partial decay width, and both the
\hfourl\ and \hgg\ branching ratios would also be affected by the changes
to the total Higgs boson decay width.
 
This section presents constraints on \kappab and \kappac,
inferred from the measured \ptH\ distributions.
A combined interpretation is then performed in terms of \kappab and
\kappac by including also the constraints from the measurement of
Higgs bosons, produced in association with a vector boson,
decaying to $b\bar{b}$~\cite{HIGG-2018-51},
and from the search for Higgs bosons produced
in a similar way and decaying to $c\bar{c}$~\cite{ATLAS:2022ers}.
All tree-level couplings of the Higgs boson to particles other than the $b$- or
$c$-quarks are set to their SM values and loop-induced Higgs boson couplings
are resolved to their SM expectation, with \kappab and \kappac as free parameters.

\subsection{Constraints from the Higgs boson transverse momentum distributions}
 
The constraints on \kappab and \kappac from the observed $\ptH$ distributions
in the \hfourl\ and \hgg\ final states are derived in two scenarios:
one in which only modification to the shape of the measured
\ptH\ distributions is considered (``{\em shape-only}''), and one
in which the impact on the overall expected normalisation, through
modifications of the total cross-sections, branching ratios and
Higgs boson decay width, is also considered (``{\em shape+normalisation}'').
 
The theoretical predictions used for these interpretations are
detailed in Ref.~\cite{ATLASgamgam}. The predictions for \kappab and
\kappac modifications of the ggF production are computed with
\scetlib~\cite{scetlib, Billis:2021ecs}.
For the \hgg\ decay channel, the calculations are performed directly
after applying the particle-level requirements defining the
\hgg\ fiducial phase space, while for the  \hfourl\ decay channel, the
calculations are performed in the full phase space and then
extrapolated to the \hfourl\ fiducial phase space using the acceptance
factors obtained from the \NNLOPS ggF prediction.
It has been verified
that the dependence of the acceptance factors in each \ptH\ bin
on the $b$ and $c$ Yukawa coupling modifiers is negligible.
The predictions
for quark-initiated \(b\bar{b}\to H\) and \(c\bar{c}\to H\) production
modes are computed with \MGNLO[2.7.3].
The simulation of the Higgs boson decay, the parton shower,
hadronisation and underlying event,
is performed with \PYTHIA[8] using a
dedicated PDF set~\cite{Bonvini:2016fgf} and the A14
tune.
The inclusive \(b\bar{b}\to H\) and \(c\bar{c}\to H\)
cross-sections are then normalised to the state-of-the-art
\NNLO\ computations available in
Refs.~\cite{Bonvini:2016fgf,Harlander_2016}. All the other Higgs boson
production modes remain unchanged with \kappab and \kappac variations,
and they are estimated as detailed in Section \ref{sec:MC}.
 
Theoretical uncertainties related to the QCD modelling and PDF
uncertainties on the differential cross-sections are considered using
the procedure detailed in Ref.~\cite{ATLASgamgam}. For the theoretical
calculation of the ggF process, uncertainties related to numerical
integration, fixed-order scale, hard resummation phase, resummation
scheme, matching scale and non-perturbative scheme are implemented
\cite{scetlib}. For the $b$- and $c$-quark initiated processes, the uncertainty
related to missing higher order QCD effects are estimated by varying
the renormalisation, factorisation and merging scales; the
uncertainties related to the PDF set are estimated by varying the mass
and scale associated with the $b$-quark for the \(b\bar{b}\to H\)
process and by using the MC replicas of the nominal PDF set for the
\(c\bar{c}\to H\) process; the uncertainties due to the choice of the
FxFx merging scale are estimated by using alternative values of this
scale.
Theoretical uncertainties in the other production modes that do not
depend on \kappab\ or \kappac, from higher-order QCD effects, PDF and
$\alpha_s$, and the parton shower model, are estimated as described in
Section~\ref{sec:MC}.
 
The statistical interpretation is performed by first parameterising
the fiducial cross-sections as a function of \kappab and \kappac for
each decay channel.
The two likelihood models are then jointly interpreted using the
same procedure as detailed in Section~\ref{sec:stat}.

\begin{table*}[!htbp]
\centering
\caption{
Observed and expected
95\% confidence intervals for the Yukawa coupling modifiers
when modifications to only the \ptH~shape are considered ({\em shape-only}),
for the individual decay channels and their combination.
The results for one coupling modifier are obtained while fixing the other
one to the SM expectation ($\kappa=1$).
}
\label{tab:interp_LY_results_shapeonly_otherkappaSM}
 
\begin{tabular}{ l c c | c }
\toprule
\multirow{2}{*}{Channel} & \multirow{2}{*}{Parameter} &  Observed                 & Expected \\
&                            &  95\% confidence interval & 95\% confidence interval\\
\midrule
\multirow{2}{*}{\hfourl}
& $\kappa_{b}$ & $[-1.8,~6.4]$  & $[-3.3,~9.3]$\\ [2pt]
& $\kappa_{c}$ & $[-7.7,~18.3]$ & $[-12.3,~19.2]$\\ [2pt]
 
\multirow{2}{*}{\hgg}
& $\kappa_{b}$  & $[-3.5,~10.2]$  & $[-2.5,~8.0]$ \\ [2pt]
& $\kappa_{c}$ & $[-12.6,~18.3]$ & $[-10.1,~17.3]$ \\ [1pt]
 
\midrule
\multirow{2}{*}{Combined}
& $\kappa_{b}$ & $[-2.0,~7.4]$   & $[-2.0,~7.4]$\\ [2pt]
& $\kappa_{c}$ & $[-8.6,~17.3]$ & $[-8.5,~15.9]$\\ [1pt]
\bottomrule
\end{tabular}
\end{table*}

The expected and observed 95\% confidence intervals
for \kappab and \kappac are shown in
Tables~\ref{tab:interp_LY_results_shapeonly_otherkappaSM}
and~\ref{tab:interp_LY_results_shapenorm_otherkappaSM} for the
{\em shape-only} and {\em shape+normalisation} scenarios, respectively.
The limits on a given
\(\kappa\) parameter are determined with
the other one fixed to SM prediction ($\kappa=1$).
If $\kappa_b$ is unconstrained in the fit, the 95\% confidence intervals
for $\kappac$ are about 10\% (twice) larger than if $\kappa_b$ is fixed
to the SM value of one, in the {\em shape-only}
({\em shape+normalisation}) approach.
 
In the {\em shape-only} approach,
the combined observed limits are less stringent than the individual
\hfourl\ result.
This is due to the quadratic dependency of the cross-section and the
differential distribution on the $\kappa$ parameters leading to a
double minimum in the profile likelihood ratio, and due to the
combined best-fit value for the \kappab parameter being further from
the SM expectation when probing only the \hgg\ decay channel.
For \kappac, the observed combined best-fit value is similar to the
best-fit value in the \hfourl\ channel. However, due to the correlation
between the \kappab and \kappac parameters, different best-fit \kappab
observations between the channels, as well as the data fluctuations in some
of the \ptH\ bins, the 68\% CL observed combined limits on \kappac are worse
than the results from the \hfourl\ channel. The corresponding 95\% CL
limits are similar to those from the \hfourl\ channel.
 
In the {\em shape+normalisation} scenario the
constraints on the coupling modifiers are
tighter, since a large fraction of the allowed
ranges for $\kappa_b$ and $\kappa_c$ from the
{\em shape-only} approach lead to values of the
total width and thus of the \hfourl\ and \hgg\
branching ratios and overall normalisation
that are inconsistent with the data.
 
Two-dimensional confidence regions on \kappab and \kappac
are also derived for both scenarios,
as shown in Figure~\ref{fig:interp_kbkc}.

\begin{table*}[htbb!]
\centering
\caption{
Observed and expected
95\% confidence intervals for the Yukawa coupling modifiers
when modifications to both the \ptH~shape and normalisation are considered ({\em shape+normalisation}),
for the individual decay channels and their combination.
The results for one coupling modifier are obtained while fixing the other
one to the SM expectation ($\kappa=1$).
}
\label{tab:interp_LY_results_shapenorm_otherkappaSM}
 
\begin{tabular}{ l c c | c }
\toprule
\multirow{2}{*}{Channel} & \multirow{2}{*}{Parameter} &  Observed                 & Expected \\
&                            &  95\% confidence interval & 95\% confidence interval\\
\midrule
\multirow{2}{*}{\hfourl}
& $\kappa_{b}$ & $[-1.14,-0.88] \cup [0.80, 1.17]$  & $[-1.23,-0.87] \cup [0.82, 1.20]$\\ [2pt]
& $\kappa_{c}$ & $[-2.94,~2.99]$                    & $[-3.33,3.14]$\\ [2pt]
 
\multirow{2}{*}{\hgg}
& $\kappa_{b}$ & $[-1.12,-0.78] \cup [0.78,1.07]$  & $[-1.18,-0.87] \cup [0.83,1.19]$ \\ [2pt]
& $\kappa_{c}$ & $[-2.46,~2.32]$                   & $[-3.03,~3.09]$ \\ [1pt]
 
\midrule
\multirow{2}{*}{Combined}
& $\kappa_{b}$ & $[-1.09,-0.86] \cup [0.81,1.09]$   & $[-1.14,-0.92] \cup [0.86,1.15]$\\ [2pt]
& $\kappa_{c}$ & $[-2.27,~2.27]$                    & $[-2.77,~2.75]$\\ [1pt]
\bottomrule
\end{tabular}
\end{table*}

\begin{figure}[!htbp]
\centering
\subfloat[]{\includegraphics[width=0.47\linewidth]{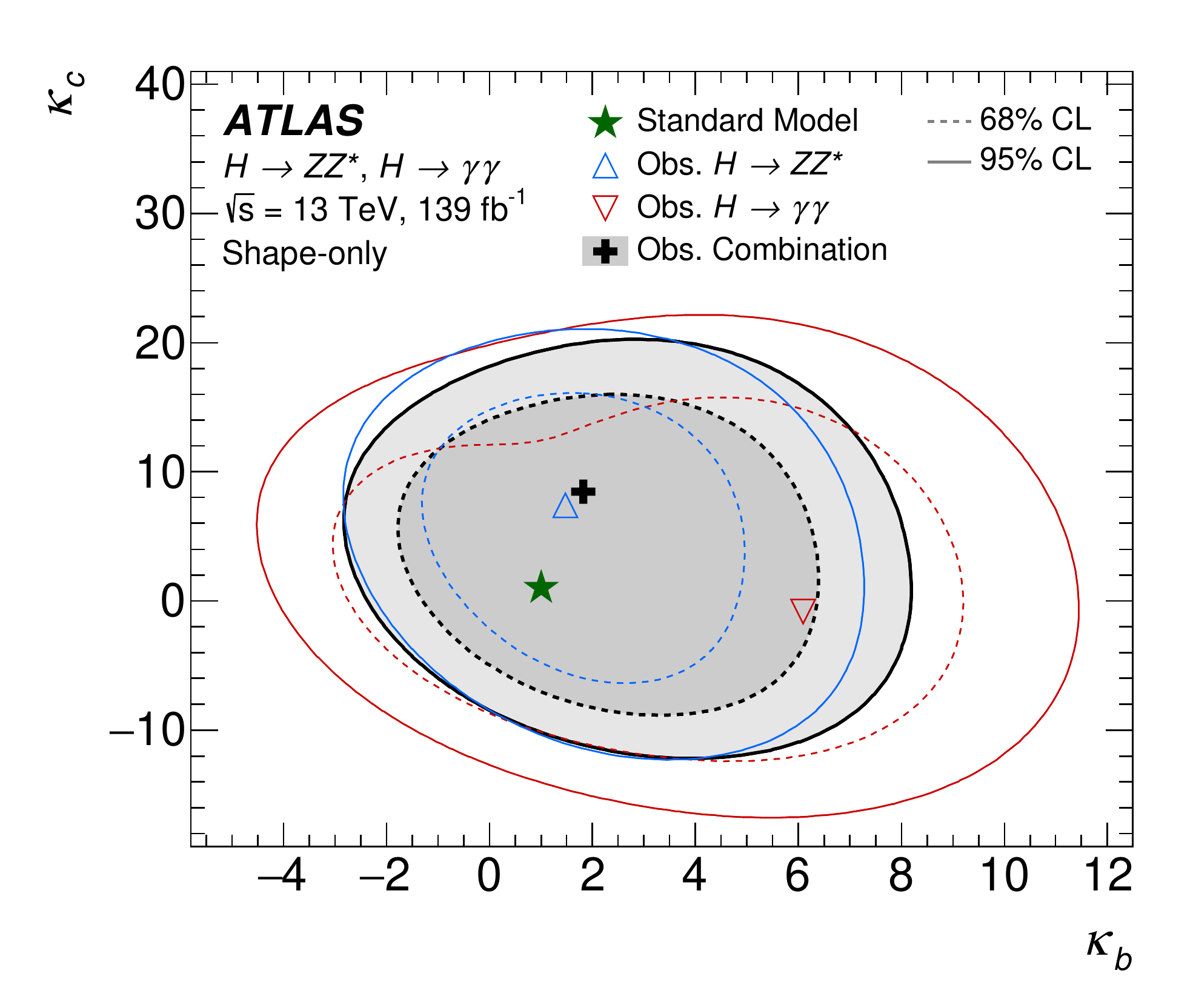}\label{fig:obs_kappa_2D_shapeonly}}
\subfloat[]{\includegraphics[width=0.47\linewidth]{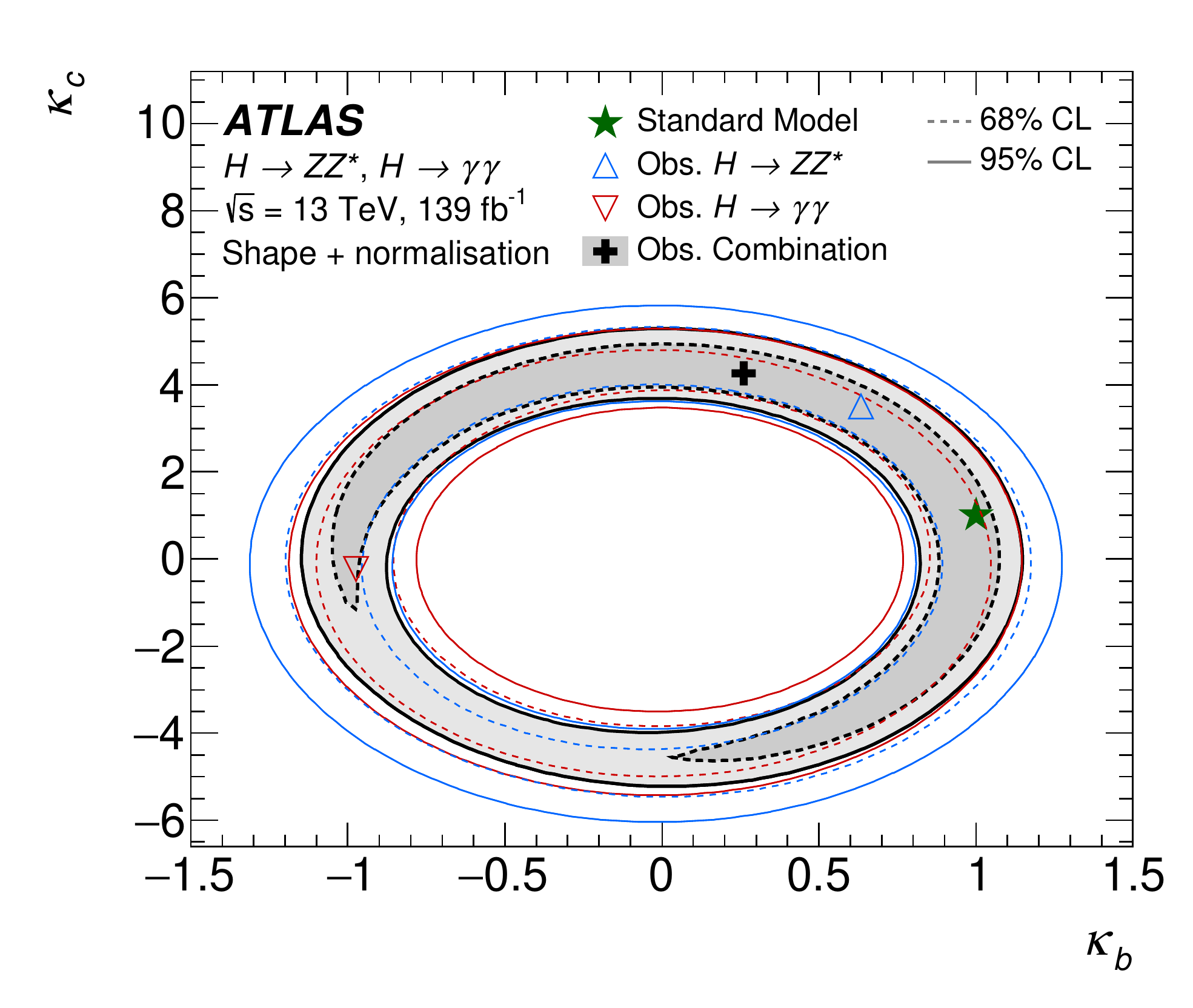}\label{fig:obs_kappa_2D_shapenorm}}
\caption{Observed limits at 95\% CL on the Yukawa coupling modifiers
\kappab and \kappac when
\protect\subref{fig:obs_kappa_2D_shapeonly} only the shape of
the \ptH\ differential cross-section ({\em shape-only})
or \protect\subref{fig:obs_kappa_2D_shapenorm} also its normalisation ({\em shape+normalisation})
is used to
constrain the parameters for the combined and individual decay
channels results. The SM predictions ($\ast$) and the observed
best-fit values (+) are indicated on the
plots.}
\label{fig:interp_kbkc}
\end{figure}
 
\FloatBarrier
 
\subsection{Combination with the constraints from $VH(b\bar{b})$ and
$VH(c\bar{c})$ production}
 
The measurement of Higgs boson decays to $b\bar{b}$ and the search for Higgs
boson decays to $c\bar{c}$ in Higgsstrahlung events ($\VH$)
constrain the $b$- and $c$-quark coupling modifiers through the
quadratic dependence on $\kappab^2$  and $\kappac^2$ of the partial widths of
the Higgs boson to these two final states.
This section describes the methodology and the results of a
simultaneous fit to the Higgs boson transverse
momentum distributions of the \hfourl\ and \hgg\ fiducial cross section
measurements and to the multivariate discriminant used to measure
the $\VH(q\bar{q})$ ($q=b,c$) signal strength~\cite{HIGG-2018-51,ATLAS:2022ers}.

Two scenarios are considered for this combination. The first scenario is the
``{\em shape+normalisation}'' scenario as described previously.
In the second scenario, the Higgs boson is also
allowed to decay to BSM particles and the associated partial width is
included in the total width.
The partial width for BSM decays is parameterised as
$\Gamma_\mathrm{BSM} = \Gamma \times B_\mathrm{BSM} = \Gamma_\mathrm{SM}\frac{B_\mathrm{BSM}}{1-B_\mathrm{BSM}}$,
where $\Gamma$ is the Higgs boson
total width, and $B_\mathrm{BSM}$ is its branching ratio to BSM particles.
The second scenario reduces the assumptions of the model, at the cost of
reduced sensitivity.
 
In the combination, most common experimental systematic uncertainties and
signal theory uncertainties are modelled as correlated between the four
channels (\hfourl, \hgg, $VH(b\bar b)$, $VH(c\bar c)$).
Jet energy calibration and flavour tagging
efficiency uncertainties are not modelled as correlated between the channels
due to the use of different jet clustering algorithms.
 
The observed 68\% and 95\% CL contours in the 2D \kappab vs \kappac plane
are shown in Figure~\ref{fig:YukawaPlusVHbbcc:contours-VHqq-bsm0} for the
{\em shape+normalisation} scenario where $B_\mathrm{BSM}$ is fixed to zero
and in Figure~\ref{fig:YukawaPlusVHbbcc:contours-VHqq-bsm} for the case
where $B_\mathrm{BSM}$ is a free parameter.
The fit prefers a positive value of \kappab, but negative values are
not excluded at 68\% CL, leading to two disconnected allowed regions,
corresponding to positive or negative values of \kappab.
One-dimensional confidence intervals for \kappac\ with
$\kappab$ unconstrained in the fit are summarised in
Table~\ref{tab:YukawaPlusVHbbcc:kc}.
Excluding the $VH(c\bar{c})$ channel would worsen the one-dimensional
constraints on $\kappa_c$ by about 10\% for the $B_\mathrm{BSM}=0$
scenario, and by a factor two for the alternative scenario
where $B_\mathrm{BSM}$ is not fixed to zero.
 
\begin{figure}[htbp]
\centering
\subfloat[]{
\includegraphics[width=0.45\textwidth]{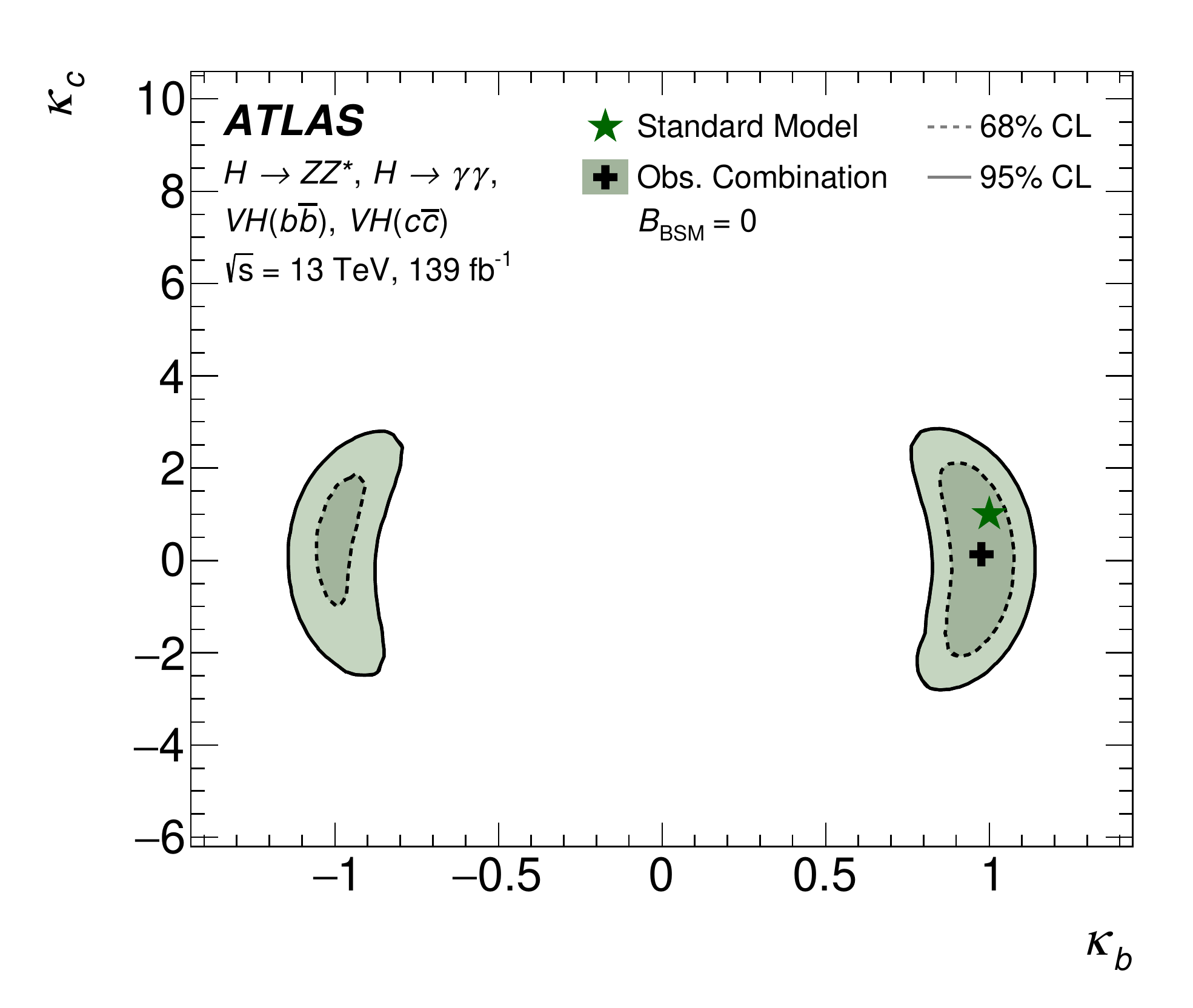}
\label{fig:YukawaPlusVHbbcc:contours-VHqq-bsm0}
}
\subfloat[]{
\includegraphics[width=0.45\textwidth]{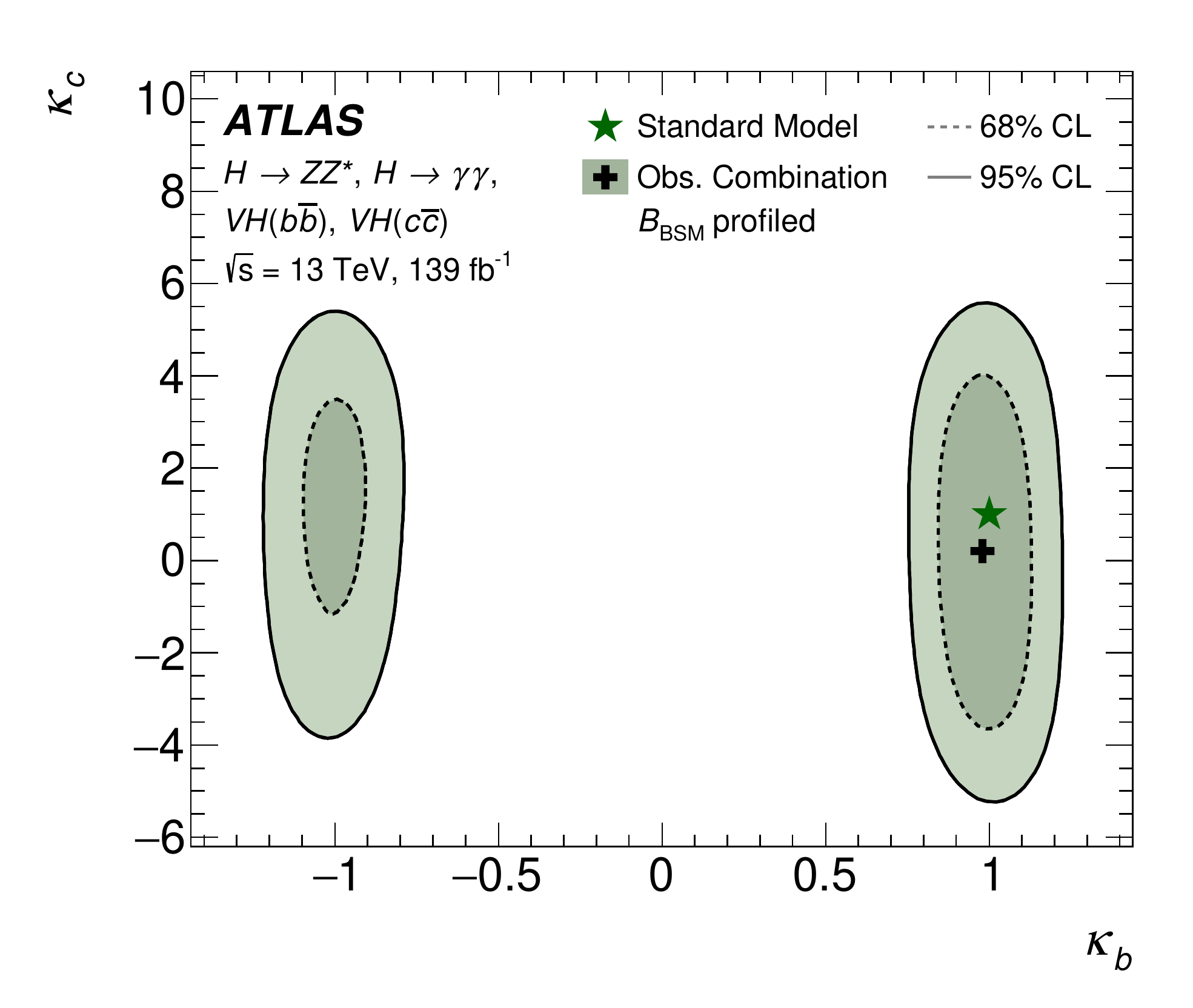}
\label{fig:YukawaPlusVHbbcc:contours-VHqq-bsm}
}
\caption{
Observed 2D negative log likelihood contours for the \kappab and \kappac
parameters from a simultaneous fit to the Higgs $\pT$
fiducial cross-sections in \hgg\ and \hfourl\ and
to multivariate discriminants used to identify
$VH$ events with Higgs bosons decaying to $b\bar{b}$ or
$c\bar{c}$, for \protect\subref{fig:YukawaPlusVHbbcc:contours-VHqq-bsm0}
$B_\mathrm{BSM}=0$ or \protect\subref{fig:YukawaPlusVHbbcc:contours-VHqq-bsm}
leaving $B_\mathrm{BSM}$ unconstrained.
}
\end{figure}

\begin{table*}[!htbp]
\centering
\caption{
One-dimensional confidence intervals in \kappac, while profiling
\kappab, at 68\% and 95\% CL, obtained from a simultaneous fit to
fiducial cross-sections in \hfourl\ and \hgg\ in bins of the Higgs
boson \pT and to $VH$ data with Higgs bosons decaying to $b\bar{b}$ or
$c\bar{c}$.
}
\label{tab:YukawaPlusVHbbcc:kc}
 
\begin{tabular}{ l c c }
\toprule
\multirow{2}{*}{Scenario} & Observed                 & Observed \\
& 68\% confidence interval & 95\% confidence interval\\
\midrule
$B_\mathrm{BSM}=0$         & $[-1.61, 1.70]$             & $[-2.47, 2.53]$\\ [3pt]
No assumption on $B_\mathrm{BSM}$  & $[-2.63, 3.01]$             & $[-4.46, 4.81]$\\ [1pt]
\bottomrule
\end{tabular}
\end{table*}

\FloatBarrier


\section{Conclusions}
A combined measurement of the total and differential Higgs production
cross-sections in the  \hgg\ and \hfourl\ decay channels was performed
using 139 \ifb\ of 13 \TeV\ proton–proton collision
recorded by the ATLAS detector during the LHC Run 2.
Good agreement is observed when comparing the results from the two
channels, after extrapolation to a common phase space.
The total Higgs boson production cross-section is measured with an
unprecedented precision of 7\%, comparable to that of the best available
Standard Model prediction which is 5\%. The result, 55.5$^{+4.0}_{-3.8}$~pb,
agrees with the SM predicted value of \hbox{55.6~$\pm$~2.5~pb}.
 
Differential cross-sections are measured as a function of the Higgs boson
transverse momentum and rapidity, the number of jets produced together
with the Higgs boson and the transverse momentum of the leading jet.
The larger data set and the combination of the
two decay channels result in measurement uncertainties that are
significantly smaller than in previous results.
Notably, the differential cross-section as a function of
the Higgs boson transverse momentum is measured with 20--30\%
precision up to 300~GeV and about 60\% precision in the
300--650~GeV range.
The combined differential distributions agree with the Standard Model
predictions.
 
The measured fiducial differential cross-sections as a function of
\ptH\ are used to derive limits on the bottom and
charm-quark Yukawa couplings modifiers, \kappab and \kappac,
assuming SM values of the other tree-level Higgs boson couplings.
Fixing the value of $\kappab$ to one,
the 95\% confidence interval for \kappac is $[-8.6, 17.3]$
using only the observed shape of the \ptH\ distribution,
and $[-2.27, 2.27]$ when considering also the impact of these couplings
on the normalisation
of the measured \ptH\ fiducial cross-sections.
 
A combined fit with the ATLAS measurement of
Higgs bosons produced in association with a $W$ or $Z$ boson and
decaying to $b$- or $c$-quark pairs allows constraints
to be set on the charm quark coupling modifier without any
assumption on the bottom quark coupling.
The 95\% CL allowed range for \kappac when the Higgs boson
is assumed to decay only to SM particles is $[-2.47, 2.53]$
while in a more generic scenario in which BSM Higgs boson
decays are allowed, the constraint is loosened to $[-4.46,4.81]$.
These represent the most stringent constraints on \kappac
to date in these scenarios.


\clearpage
 
\section*{Acknowledgements}


We thank CERN for the very successful operation of the LHC, as well as the
support staff from our institutions without whom ATLAS could not be
operated efficiently.
 
We acknowledge the support of
ANPCyT, Argentina;
YerPhI, Armenia;
ARC, Australia;
BMWFW and FWF, Austria;
ANAS, Azerbaijan;
CNPq and FAPESP, Brazil;
NSERC, NRC and CFI, Canada;
CERN;
ANID, Chile;
CAS, MOST and NSFC, China;
Minciencias, Colombia;
MEYS CR, Czech Republic;
DNRF and DNSRC, Denmark;
IN2P3-CNRS and CEA-DRF/IRFU, France;
SRNSFG, Georgia;
BMBF, HGF and MPG, Germany;
GSRI, Greece;
RGC and Hong Kong SAR, China;
ISF and Benoziyo Center, Israel;
INFN, Italy;
MEXT and JSPS, Japan;
CNRST, Morocco;
NWO, Netherlands;
RCN, Norway;
MEiN, Poland;
FCT, Portugal;
MNE/IFA, Romania;
MESTD, Serbia;
MSSR, Slovakia;
ARRS and MIZ\v{S}, Slovenia;
DSI/NRF, South Africa;
MICINN, Spain;
SRC and Wallenberg Foundation, Sweden;
SERI, SNSF and Cantons of Bern and Geneva, Switzerland;
MOST, Taiwan;
TENMAK, T\"urkiye;
STFC, United Kingdom;
DOE and NSF, United States of America.
In addition, individual groups and members have received support from
BCKDF, CANARIE, Compute Canada and CRC, Canada;
PRIMUS 21/SCI/017 and UNCE SCI/013, Czech Republic;
COST, ERC, ERDF, Horizon 2020 and Marie Sk{\l}odowska-Curie Actions, European Union;
Investissements d'Avenir Labex, Investissements d'Avenir Idex and ANR, France;
DFG and AvH Foundation, Germany;
Herakleitos, Thales and Aristeia programmes co-financed by EU-ESF and the Greek NSRF, Greece;
BSF-NSF and MINERVA, Israel;
Norwegian Financial Mechanism 2014-2021, Norway;
NCN and NAWA, Poland;
La Caixa Banking Foundation, CERCA Programme Generalitat de Catalunya and PROMETEO and GenT Programmes Generalitat Valenciana, Spain;
G\"{o}ran Gustafssons Stiftelse, Sweden;
The Royal Society and Leverhulme Trust, United Kingdom.
 
The crucial computing support from all WLCG partners is acknowledged gratefully, in particular from CERN, the ATLAS Tier-1 facilities at TRIUMF (Canada), NDGF (Denmark, Norway, Sweden), CC-IN2P3 (France), KIT/GridKA (Germany), INFN-CNAF (Italy), NL-T1 (Netherlands), PIC (Spain), ASGC (Taiwan), RAL (UK) and BNL (USA), the Tier-2 facilities worldwide and large non-WLCG resource providers. Major contributors of computing resources are listed in Ref.~\cite{ATL-SOFT-PUB-2021-003}.


\clearpage
\appendix
\part*{Appendix}
\addcontentsline{toc}{part}{Appendix}

\section{Correlation matrices between the measured cross-sections}
\label{app:corrmat}
 
Figure~\ref{fig:diffxsCorr} and Figure \ref{fig:diffxsCorr2} show the
correlation matrices among the differential cross-sections measured
in different bins of the same one-dimensional measurement.
 
\begin{figure}[!htbp]
\centering
\subfloat[]{\includegraphics[width=0.48\linewidth]{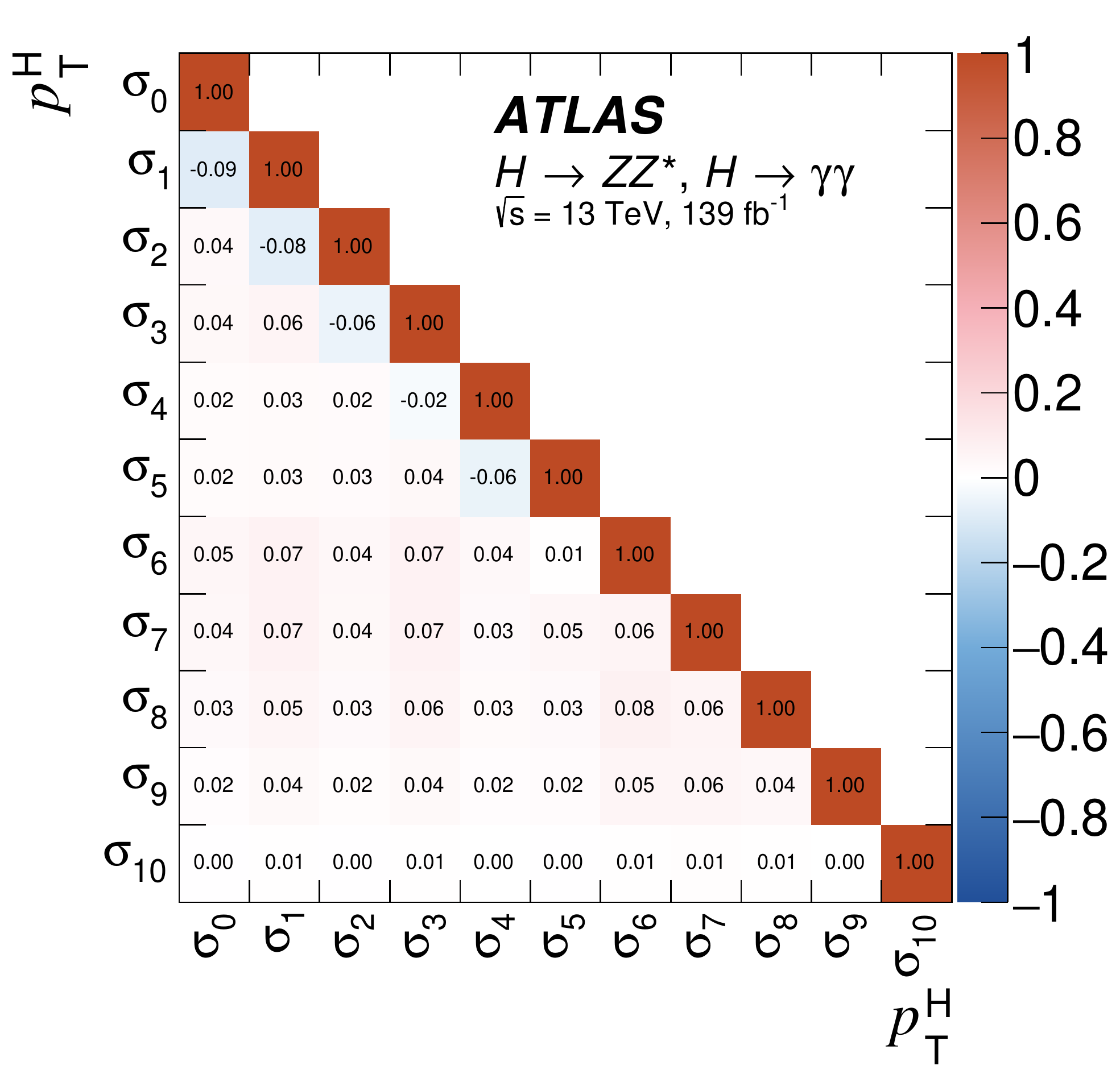}\label{fig:corrMatrix_pt}}
\subfloat[]{\includegraphics[width=0.48\linewidth]{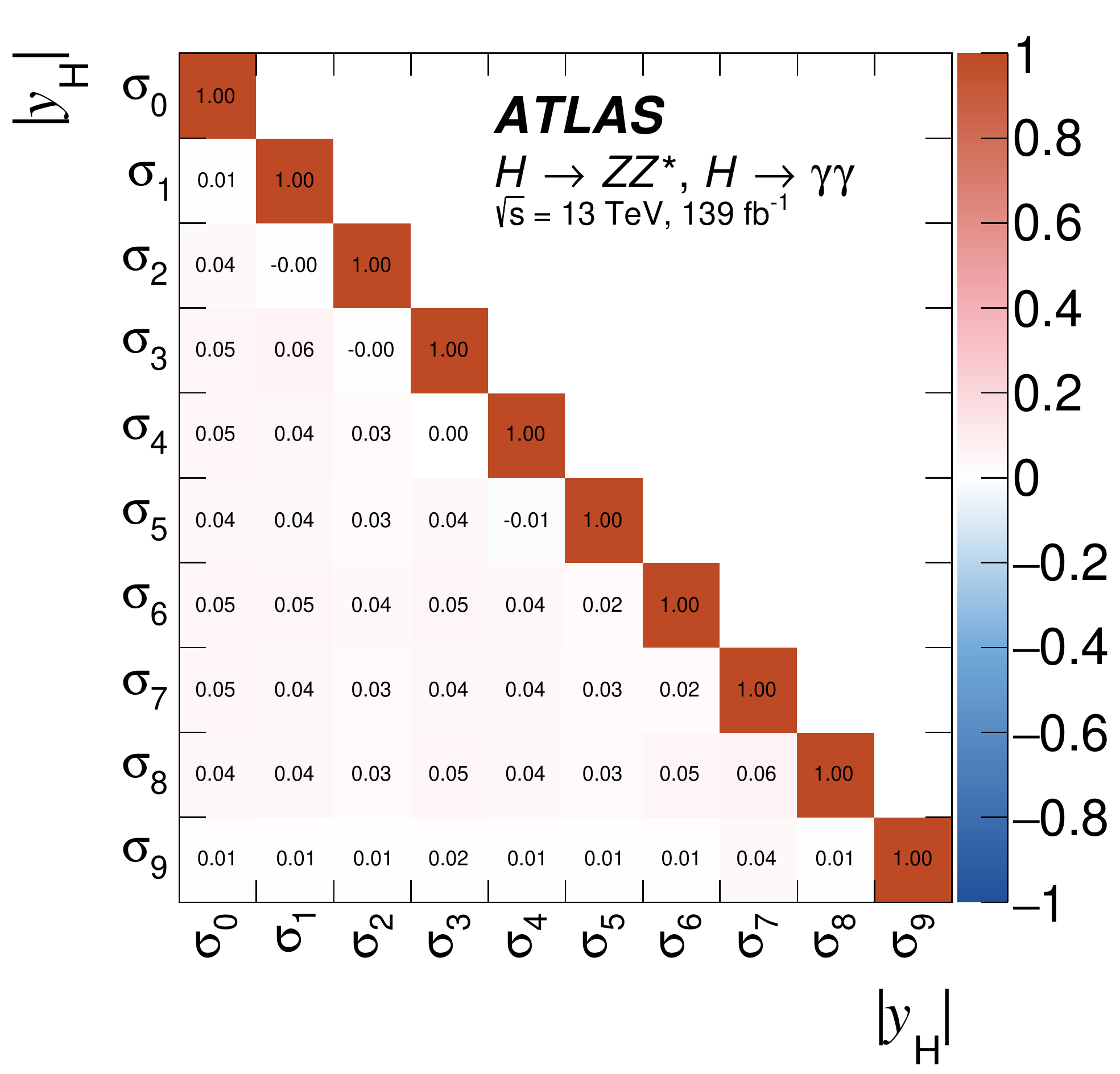}\label{fig:corrMatrix_y}}\\
\subfloat[]{\includegraphics[width=0.48\linewidth]{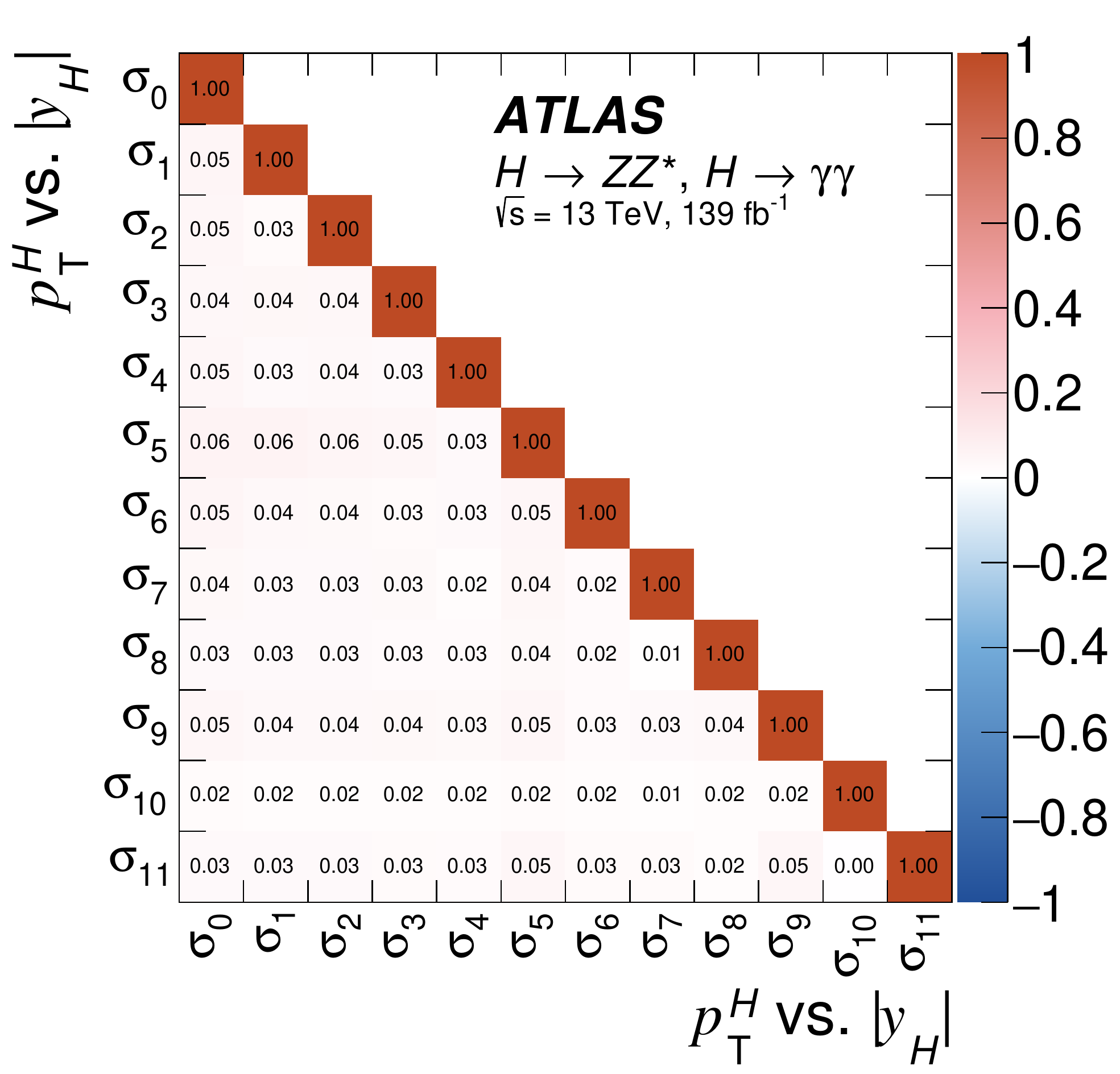}\label{fig:corrMatrix_pt_vs_y}}
\caption{Correlation matrices between the differential
$pp \rightarrow H + X$ cross-sections measured in different bins
of the same observable:
\protect\subref{fig:corrMatrix_pt} Higgs boson transverse
momentum, \protect\subref{fig:corrMatrix_y} Higgs boson rapidity and
\protect\subref{fig:corrMatrix_pt_vs_y} Higgs boson transverse
momentum vs Higgs boson rapidity. The labels are defined as per the bin boundaries outlined in Table \ref{tab:BinDefinition},
with a higher label index corresponding to a higher bin for the given variable. For the correlation matrix for the Higgs boson transverse
momentum vs Higgs boson rapidity, lower rapidity bins are labelled first with ascending bins in \ptH.  }
\label{fig:diffxsCorr}
\end{figure}
 
\begin{figure}[!htbp]
\centering
\centering
\subfloat[]{\includegraphics[width=0.48\linewidth]{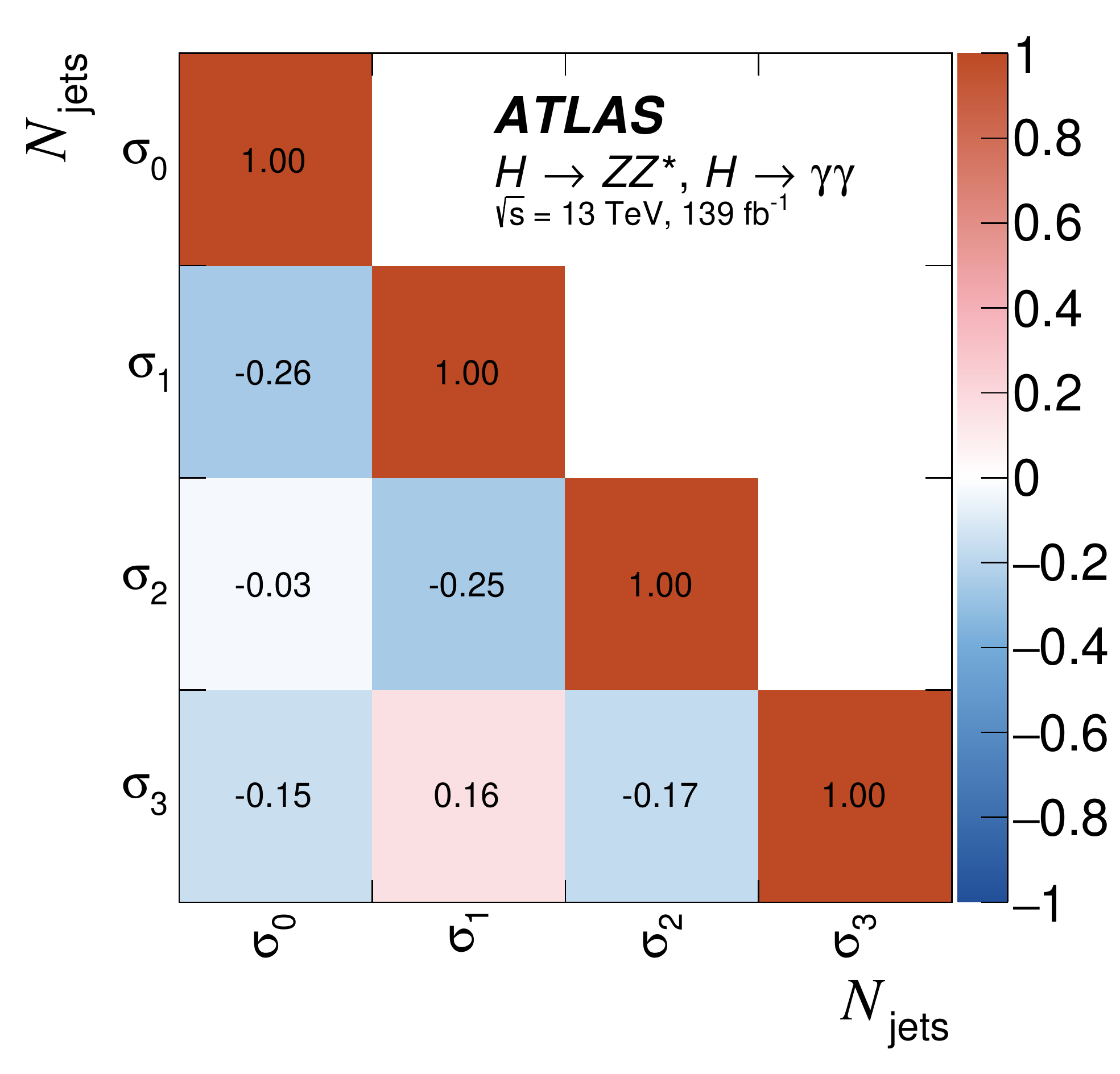}\label{fig:corrMatrix_Njet}}
\subfloat[]{\includegraphics[width=0.48\linewidth]{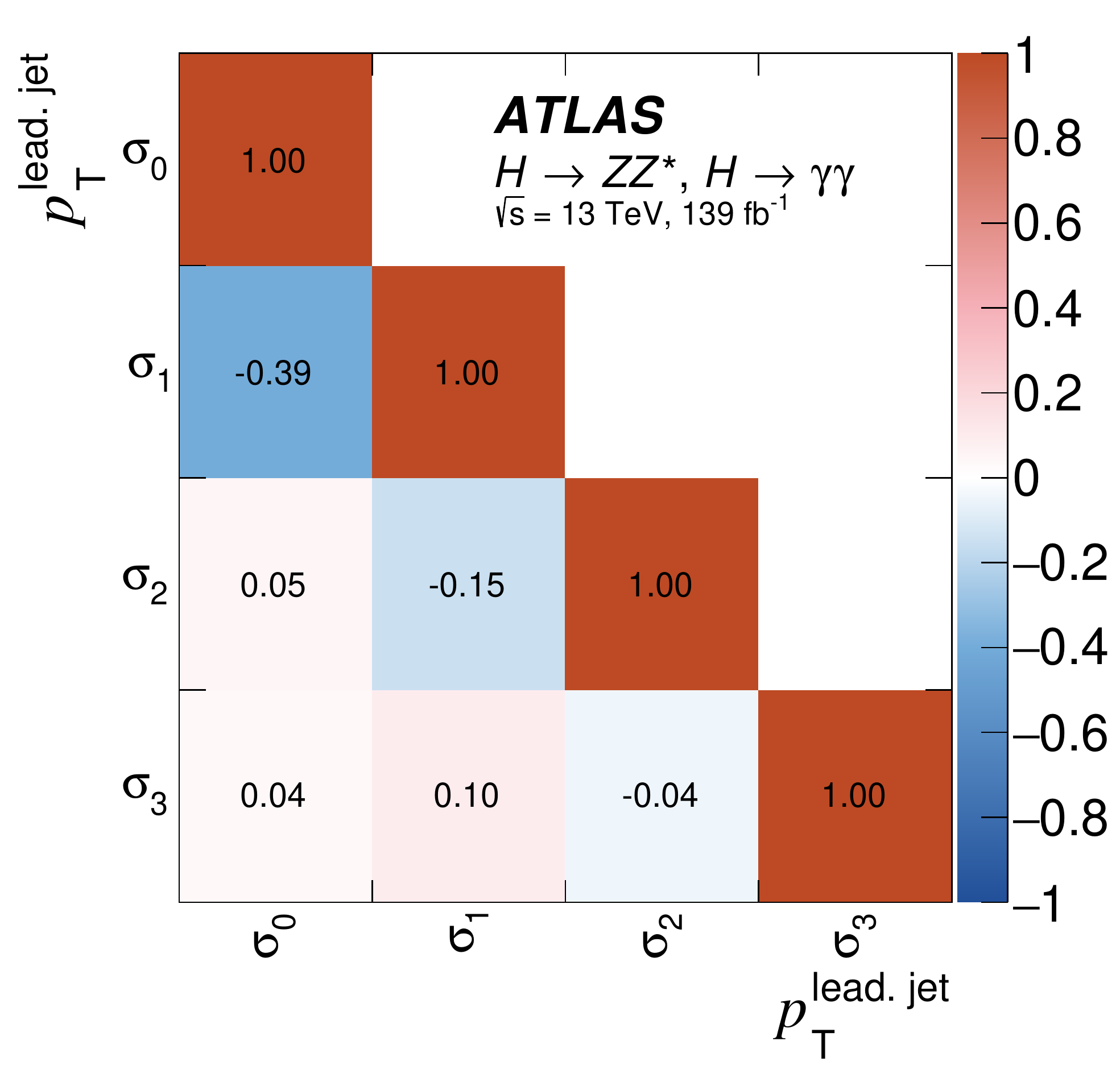}\label{fig:corrMatrix_leadjet}}
\caption{Correlation matrices between the differential $pp \rightarrow H + X$ cross-sections measured in different bins of the same observable:
\protect\subref{fig:corrMatrix_Njet} number of jets and
\protect\subref{fig:corrMatrix_leadjet} $\pT$ of the leading jet. The labels are defined as per the bin boundaries outlined in
Table \ref{tab:BinDefinition}, with a higher label index corresponding to a higher bin for the given variable.}
\label{fig:diffxsCorr2}
\end{figure}

\FloatBarrier


\printbibliography
 
\clearpage
 
\begin{flushleft}
\hypersetup{urlcolor=black}
{\Large The ATLAS Collaboration}

\bigskip

\AtlasOrcid[0000-0002-6665-4934]{G.~Aad}$^\textrm{\scriptsize 101}$,
\AtlasOrcid[0000-0002-5888-2734]{B.~Abbott}$^\textrm{\scriptsize 119}$,
\AtlasOrcid[0000-0002-7248-3203]{D.C.~Abbott}$^\textrm{\scriptsize 102}$,
\AtlasOrcid[0000-0002-1002-1652]{K.~Abeling}$^\textrm{\scriptsize 55}$,
\AtlasOrcid[0000-0002-8496-9294]{S.H.~Abidi}$^\textrm{\scriptsize 29}$,
\AtlasOrcid[0000-0002-9987-2292]{A.~Aboulhorma}$^\textrm{\scriptsize 35e}$,
\AtlasOrcid[0000-0001-5329-6640]{H.~Abramowicz}$^\textrm{\scriptsize 150}$,
\AtlasOrcid[0000-0002-1599-2896]{H.~Abreu}$^\textrm{\scriptsize 149}$,
\AtlasOrcid[0000-0003-0403-3697]{Y.~Abulaiti}$^\textrm{\scriptsize 116}$,
\AtlasOrcid[0000-0003-0762-7204]{A.C.~Abusleme~Hoffman}$^\textrm{\scriptsize 136a}$
% ...
\bigskip
\\

% ...
$^{29}$Physics Department, Brookhaven National Laboratory, Upton NY; United States of America.\\
% ...
$^{35}$$^{(a)}$Facult\'e des Sciences Ain Chock, R\'eseau Universitaire de Physique des Hautes Energies - Universit\'e Hassan II, Casablanca;$^{(b)}$Facult\'{e} des Sciences, Universit\'{e} Ibn-Tofail, K\'{e}nitra;$^{(c)}$Facult\'e des Sciences Semlalia, Universit\'e Cadi Ayyad, LPHEA-Marrakech;$^{(d)}$LPMR, Facult\'e des Sciences, Universit\'e Mohamed Premier, Oujda;$^{(e)}$Facult\'e des sciences, Universit\'e Mohammed V, Rabat;$^{(f)}$Institute of Applied Physics, Mohammed VI Polytechnic University, Ben Guerir; Morocco.\\
% ...
$^{55}$II. Physikalisches Institut, Georg-August-Universit\"{a}t G\"ottingen, G\"ottingen; Germany.\\
% ...
$^{101}$CPPM, Aix-Marseille Universit\'e, CNRS/IN2P3, Marseille; France.\\
$^{102}$Department of Physics, University of Massachusetts, Amherst MA; United States of America.\\
% ...
$^{116}$Department of Physics, New York University, New York NY; United States of America.\\
% ...
$^{119}$Homer L. Dodge Department of Physics and Astronomy, University of Oklahoma, Norman OK; United States of America.\\
% ...
$^{136}$$^{(a)}$Departamento de F\'isica, Pontificia Universidad Cat\'olica de Chile, Santiago;$^{(b)}$Millennium Institute for Subatomic physics at high energy frontier (SAPHIR), Santiago;$^{(c)}$Instituto de Investigaci\'on Multidisciplinario en Ciencia y Tecnolog\'ia, y Departamento de F\'isica, Universidad de La Serena;$^{(d)}$Universidad Andres Bello, Department of Physics, Santiago;$^{(e)}$Instituto de Alta Investigaci\'on, Universidad de Tarapac\'a, Arica;$^{(f)}$Departamento de F\'isica, Universidad T\'ecnica Federico Santa Mar\'ia, Valpara\'iso; Chile.\\
% ...
$^{149}$Department of Physics, Technion, Israel Institute of Technology, Haifa; Israel.\\
$^{150}$Raymond and Beverly Sackler School of Physics and Astronomy, Tel Aviv University, Tel Aviv; Israel.\\
% ...
$^{a}$ Also Affiliated with an institute covered by a cooperation agreement with CERN.\\
$^{b}$ Also at Borough of Manhattan Community College, City University of New York, New York NY; United States of America.\\
$^{c}$ Also at Bruno Kessler Foundation, Trento; Italy.\\
$^{d}$ Also at Center for High Energy Physics, Peking University; China.\\
$^{e}$ Also at Center for Interdisciplinary Research and Innovation (CIRI-AUTH), Thessaloniki ; Greece.\\
$^{f}$ Also at Centro Studi e Ricerche Enrico Fermi; Italy.\\
% ...

\end{flushleft}


\end{document}